\documentclass[ twocolumn, tightenlines, twoside, secnumarabic,
superscriptaddress, preprintnumbers, nofootinbib, notitlepage]{revtex4-1}

\usepackage[colorlinks=true,citecolor=blue,linkcolor=blue]{hyperref}
\usepackage[normalem]{ulem}
\usepackage{amsmath,amssymb}
\usepackage{epsfig}
\usepackage{graphicx}        
\usepackage{url}
\usepackage{color}
\usepackage{multirow}
\usepackage{placeins}
\usepackage[dvipsnames]{xcolor}
\usepackage{epstopdf}
\usepackage{tikz}
\usetikzlibrary{trees}
\usetikzlibrary{decorations.pathmorphing}
\usetikzlibrary{decorations.markings}

\definecolor{jblue} {RGB}{20,50,100}
\definecolor{npurple} {RGB} {153, 51, 204}
\definecolor{wred}  {RGB}{217,0,56}
\definecolor{white}  {RGB}{255,255,255}

\definecolor{korange}  {RGB}{235, 80, 43}
\definecolor{korange2}  {RGB}{245, 100, 63}
\definecolor{kyelloworange}  {RGB}{255, 210, 110}
\definecolor{kyelloworange2}  {RGB}{240, 170, 90}
\definecolor{kred}  {RGB}{204, 102, 153}
\definecolor{kpurple}  {RGB}{153, 61, 190}
\definecolor{kpurplelight}  {RGB}{213, 161, 230}

\usepackage{cleveref}
\definecolor{red}{rgb}{1.0, 0, 0}
\allowdisplaybreaks

\bibpunct{[}{]}{,}{n}{}{,}

\renewcommand{\vec}[1]{{\mathbf{#1}}}


\newcommand{\nn}{\nonumber}

\begin{document}

\title{A Particle Physics Origin of the 5 MeV Bump in the Reactor Antineutrino Spectrum?}

\author{Jeffrey M. Berryman}
\affiliation{Center for Neutrino Physics, Department of Physics, Virginia Tech, Blacksburg, VA 24061, USA}
\author{Vedran Brdar}
\affiliation{Max-Planck-Institut f\"{u}r Kernphysik, 69117 Heidelberg, Germany}
\author{Patrick Huber}
\affiliation{Center for Neutrino Physics, Department of Physics, Virginia Tech, Blacksburg, VA 24061, USA}

\begin{abstract}
One of the most puzzling questions in neutrino physics is the origin
of the excess at 5 MeV in the reactor antineutrino spectrum.  In this
paper, we explore the excess via the reaction
$^{13}$C$(\overline{\nu}, \overline{\nu}^\prime n)^{12}$C$^*$ in
organic scintillator detectors. The de-excitation of $^{12}$C$^*$
yields a prompt $4.4$\,MeV photon, while the thermalization of the
product neutron causes proton recoils, which in turn yield an
additional prompt energy contribution with finite width. Together,
these effects can mimic an inverse beta decay event with around 5\,MeV
energy. We consider several non-standard neutrino interactions to
produce such a process and find that the parameter space preferred by
Daya Bay is disfavored by measurements of neutrino-induced deuteron
disintegration and coherent elastic neutrino-nucleus scattering. While
non-minimal particle physics scenarios may be viable, a nuclear
physics solution to this anomaly appears more appealing.

\end{abstract}

\maketitle

\section{Introduction}
\label{sec:intro}
Neutrinos\footnote{For brevity we will use the term neutrino for both
  neutrinos and antineutrinos.} from nuclear reactors have a prominent
role in the field of neutrino physics; neutrinos were discovered using
reactors as sources, and $\theta_{13}$ was shown to be nonzero using
detectors close to a nuclear reactor
\cite{Abe:2011fz,An:2012eh,Ahn:2012nd}. Currently, reactor experiments
are grappling with two anomalous findings. The first is the $\sim 6$\%
mismatch between the predicted and observed rates in short-baseline
reactor experiments that could be explained via oscillations into
additional, eV-scale neutrinos \cite{Mention:2011rk}. The other, more
recent one is the excess of neutrino events around 5\,MeV, the
so-called reactor bump \cite{Huber:2016fkt}. This spectral feature has
triggered significant interest in the community in recent years
\cite{Huber:2016xis,Buck:2015clx,Dwyer:2014eka}, but has thus far
eluded any quantitative explanation in terms of nuclear
physics. Moreover, no attempts in terms of beyond the Standard Model
(SM) physics can be found in the literature. Due to the lack of
general agreement on the source of the discrepancy, it appears
meaningful to explore potential for explaining the 5 MeV bump with new
physics (NP).

Current reactor experiments detect neutrinos via inverse beta decay
(IBD): $\bar\nu_e +p\rightarrow n + e^+$. The signature used to
identify these events and separate them from background is the delayed
coincidence between the positron, $e^+$ -- which forms the prompt
signal -- and the neutron capture on gadolinium and the subsequent
emission of a gamma cascade of about 8\,MeV total energy -- the delayed
signal, some 50--100\,$\mu$s later. The positron signal has no specific
characteristics and any prompt energy deposition in the detector could
be mistaken as the prompt signal. It is the delayed neutron capture
signal which is essentially background free, {\it i.e.}, only neutrons
can cause it, which then allows to tag the prompt signal as being part
of an IBD event. Under these conditions, it is clear that any new
physics explanation must produce a neutron in its final state. It also
has been established that the bump tracks reactor power and thus has
to be caused by a particle flux from the reactor which is closely
related to the number of beta decays inside the reactor. Beta decays
of any known isotope extend at most to Q-values of 15--20\,MeV, and thus, 
direct production of a neutron from any particle flux from a reactor
is excluded. Therefore, a neutron which is already present in the
detector material has to be liberated from its nuclear bound state.

Liquid scintillator as used in Daya Bay and NEOS is made up of carbon
and hydrogen. Hydrogen has only about a 1:10,000 admixture of
deuterium; thus, the only neutron source is carbon. Carbon-12 has a
neutron separation energy of 18.7\,MeV and thus there are no particles
energetic enough coming from the reactor which can effect this
reaction. On the other hand, carbon-13 has an abundance of about 1\%
and a neutron separation energy of only 4.9\,MeV, well within the
energies available in beta decay. The reaction $^{13}$C$(\bar{\nu},
\bar{\nu}^\prime n)^{12}$C can proceed via the
ground state of $^{12}$C or via the first excited state at 4.4\,MeV,
which then would de-excite immediately via the emission of a 4.4\,MeV
gamma, which naturally would appear as a line-like energy deposition
close to the bump energy. Additional prompt energy will be supplied by
the neutron kinetic energy: the neutron loses its energy by
collisions with carbon and hydrogen and the recoiling nuclei will
leave a scintillation signature. Due to the details of scintillation
light emission, explained in more detail later, and the stochastic
nature of neutron energy losses, this recoil signal will add the
missing energy to match the mean bump energy as well as give the
requisite width to the signal. Note, that this is a minimal scenario
from a phenomenological view point and apart from the interaction
causing the anomalously high $^{13}$C scattering rate, all ingredients
are well-known Standard Model physics. In particular, the bump
position and width are determined by known physics.

In this paper, we will first explore the details of signal formation
and demonstrate that it actually provides an excellent description of
the experimental data. Then we proceed, in the spirit of low-energy
effective field theory, to explore possible operators that could
cause the anomalously high $^{13}$C scattering rate and examine if
other data rule out these operators at the required strength.
We are looking for neutral-current-like operators of the form
\begin{equation}
  \label{eq:operator}
2\sqrt{2}G_f \epsilon (\bar\nu_s L\nu_s)(\bar N L N)\,,
\end{equation}
where $N$ denotes a nucleon and $L$ is an arbitrary Lorentz structure, 
{\it i.e.}, $L= 1, \, \gamma_5, \, \gamma^\mu, \, \gamma^\mu \gamma_5, \,
\ldots$. The strength of the new interaction, $\epsilon$, is 
bounded from below by the fact that the bump is about 10\% of
the regular IBD signal. Combined with the 1\% abundance of $^{13}$C,
this implies that $\epsilon>10$. Invariance under SU(2)$_\mathrm{EW}$ implies
the existence of a corresponding term involving charged leptons; any such
operator with the required interaction strength will be ruled out by charged-lepton 
data. We circumvent this restriction by introducing an additional neutrino 
state that is not part of an electroweak multiplet.\footnote{We refrain from 
calling this additional neutrino ``sterile,'' since we will endow it with NP interactions.} 
We can choose the new mixing angle and new mass splitting to be consistent 
with this additional neutrino being the explanation of the reactor antineutrino
anomaly~\cite{Mention:2011rk}.

We argue that this constitutes a phenomenological minimal framework to
seek a BSM explanation of the 5 MeV bump and we provide
the quantitative demonstration that this framework indeed explains the
reactor data. We also will demonstrate that any minimal incarnation of
the operator in Eq.~\ref{eq:operator} is ruled out by either
COHERENT~\cite{Akimov:2017ade} data or past reactor neutrino deuterium
scattering results~\cite{Reines:1980pc}.

\section{Cross section and Neutrino fluxes}
\label{sec:xsec+flux}

\subsection{Obtaining the new physics cross section}

A first-principles calculation of the $^{13}$C$(\bar{\nu},\bar{\nu}^\prime n)^{12}$C$^{(*)}$ cross section for a given NP model is beyond the scope of this work. Here, we illustrate our estimation of the cross section using measurements of $^{13}$C$(e,e^\prime n)^{12}$C$^{(*)}$ as a proxy for the NP $^{13}$C$(\bar{\nu},\bar{\nu}^\prime n)^{12}$C$^{(*)}$ cross section. In doing so, we are implicitly assuming the existence of a new vector interaction; we discuss this scenario in more detail in  \cref{sec:models}. We assume that nuclear physics effects in the exchange of the hidden vector particle, dubbed $X$, are not qualitatively different from the SM photon case.

The differential cross section for $^{13}$C$(e,e^\prime n)^{12}$C$^{(*)}$ is parameterized in Ref.~\cite{Suzuki:1999hr} by
\begin{equation}
\label{eq:xsec}
\frac{d^2\sigma}{dE_n d\Omega} (E_i, E_n) = 4 \pi A_0 (E_n) \frac{d\sigma_{\rm Mott}}{d \Omega_f} \left(E_i, E_n \right)\,,
\end{equation}
where $E_i$ is the initial-state electron energy, $E_n$ is the neutron kinetic energy, $\Omega_f$ is the final-state electron solid angle, $A_0$ is an experimentally-determined function of the energy transferred to the nuclear system \cite{Suzuki:1999hr} and $d\sigma_{\text{Mott}}/d \Omega_f$ is the differential Mott cross section for an electron scattering off a heavy, point-like nucleus in QED.

The Mott cross section in \cref{eq:xsec} is formally divergent; we regulate it by introducing a finite mass $M_X$ for $X$ and find
\begin{equation}
\label{eq:diffMott}
\frac{d\sigma_{\rm Mott}}{d \Omega_f} \left(E_i, E_n \right) = \frac{2 \alpha^2 Z^2 E_i E_f (1-\cos\theta_f)}{\left(M_X^2 + 2 E_i E_f (1-\cos\theta_f)\right)^2},
\end{equation}
where $E_f = E_i - E_n - E_{\rm th}$ is the final-state electron energy and $E_{\rm th}$ is the threshold energy for the reaction. For transitions to the ground state of $^{12}$C, $E_{\rm th} \approx$ 4.9 MeV, while for transitions to the first excited state, $E_{\rm th} \approx$ 9.3 MeV. Integrating \cref{eq:diffMott} over $d\Omega_f$ yields
\begin{align}
\label{eq:Mott}
\sigma_{\rm Mott} = & \, \, \frac{\pi \alpha^2 Z^2}{E_i E_f} \times \\
& \left[ \log \left( \frac{4 E_i E_f + M_X^2}{M_X^2} \right) - \frac{4 E_i E_f}{4 E_i E_f + M_X^2} \right]. \nonumber
\end{align}
We convert the $(e, e^\prime n)$ cross section $\sigma$ into the NP cross section $\sigma^\prime$ for the $(\bar{\nu}, \bar{\nu}^\prime n)$ reaction by rescaling the former by an effective coupling $\zeta^2$. The resulting neutron spectrum is
\begin{equation}
\label{eq:xsec2}
\frac{d\sigma^\prime}{dE_n}(E_{\overline{\nu}}, E_n) = {\zeta^2} \times \frac{d\sigma}{dE_n}(E_i = E_{\overline{\nu}}, E_n)\,,
\end{equation}
where $E_{\bar{\nu}}$ is the energy of an incoming neutrino and $d\sigma/dE_n = 4 \pi A_0 \sigma_{\rm Mott}$. Convolving this neutron spectrum with the reactor antineutrino flux, $\phi_{\bar{\nu}}$, yields the cross-section-weighted neutron spectrum, $\widetilde{\phi}_n$,
\begin{equation}
\label{eq:Nspec}
\frac{d\widetilde{\phi}_n}{d E_n} (E_n) = 0.006 \times \int d E_{\overline{\nu}} \, \frac{d\sigma^\prime}{dE_n}(E_{\overline{\nu}}, E_n) \cdot \frac{d \phi_{\bar{\nu}}}{d E_{\overline{\nu}}}(E_{\overline{\nu}}),
\end{equation}
where ``0.006'' accounts for the difference in the number of protons and carbon atoms in the linear alkylbenzene (LAB), $\left( \text{C}_6 \text{H}_5 \right) \text{C}_n \text{H}_{2n+1}$ ($n\sim10-16$), used by Daya Bay, as well as the natural abundance of $^{13}$C ($1.07\%$). Once detector effects are included, this spectrum is added to the IBD prompt-energy spectrum in order to replicate the 5 MeV bump.

\subsection{Reactor Antineutrino fluxes}

The $^{13}$C$(\bar{\nu}, \bar{\nu}^\prime n)^{12}$C$^*$ reaction requires a significant neutrino flux above $E_{\bar{\nu}} \gtrsim9.4$\,MeV, though reactor experiments have measured the flux only up to 8\,MeV, {\it e.g.}, see Ref.~\cite{An:2016srz}. The Huber-Mueller (HM) fluxes \cite{Huber:2011wv,Mueller:2011nm} do not extend beyond 9\,MeV. On the other hand, ab initio calculations of reactor fluxes~\cite{Fallot:2012jv} predict a nonzero flux up to energies of 16\,MeV. Preliminary upper bounds of the neutrino flux beyond 8\,MeV have been reported by Daya Bay~\cite{NRThesis} and our fit does not exceed these bounds. We augment the HM fluxes by including an additional power-law component beyond 8\,MeV, described by a power-law index $I$, such that the flux is continuous at $E_{\bar{\nu}} = 8$ MeV.

A popular alternative explanation for the reactor bump is that the neutrino flux from $^{238}$U has been miscalculated; there exists mounting evidence supporting a connection between the strength of the reactor bump and the $^{238}$U fuel fraction of the reactor (see Refs.~\cite{Hayes:2015yka,Huber:2016xis} for more details). This explanation and that proposed here are not mutually exclusive; if the flux of high-energy antineutrinos from $^{238}$U has been underestimated, then this may provide the additional flux that we require to reproduce the bump.

\subsection{Neutron thermalization}

Neutrons lose energy via elastic collisions with the atoms in the scintillator; we simulate these stochastic energy losses with a simple Monte Carlo calculation. To determine the amount of prompt energy from the proton recoils, we account for quenching in the liquid scintillator, assuming nonrelativistic protons. The amount of energy produced by the scintillator is given by Birks' law~\cite{Birks:1951boa},
\begin{equation}
\label{eq:birk}
E = \int \frac{ \left( \frac{dE}{dx} \right) dx}{1 + k_B \left( \frac{dE}{dx} \right)},
\end{equation}
where $k_B = 0.0065$ g cm$^{-2}$ MeV$^{-1}$ is Birks' constant for Gd-doped LAB~\cite{Minfeng} as used in Daya Bay. The quenching factor $Q = E(k_B)/E(k_B\to0)$ is calculated via \cref{eq:birk} using the proton $dE/dx$  in organic scintillator~\cite{pstar}. The combined effects of the stochastic proton energy distribution and quenching give rise to a finite width for the prompt energy reconstructed from this process.

\section{FIT}
\label{sec:fit}
We simulate $^{13}$C--$\overline{\nu}$ events using Monte Carlo methods for the following ranges of $X$ masses, effective couplings and power-law indices to determine the prompt energy spectrum at Daya Bay detector(s):
\begin{eqnarray}
M_X & \in & \text{[1 keV, 1 GeV]} \nonumber \\
\log_{10} \zeta & \in & [-12.3, \, 1.7] \nonumber \\
I & \in & [5, \, 20] \nonumber
\end{eqnarray}
The smeared spectrum is compared against the ratio of measurement-to-expectation of the prompt energy spectrum reported in Ref.~\cite{Huber:2016xis} using a standard $\chi^2$-test, incorporating uncertainties on and correlations between the Daya Bay data, theoretical uncertainties in the HM fluxes and experimental uncertainties on $A_0(E_n)$.

\begin{figure}[!tbp]
\includegraphics[width=\linewidth]{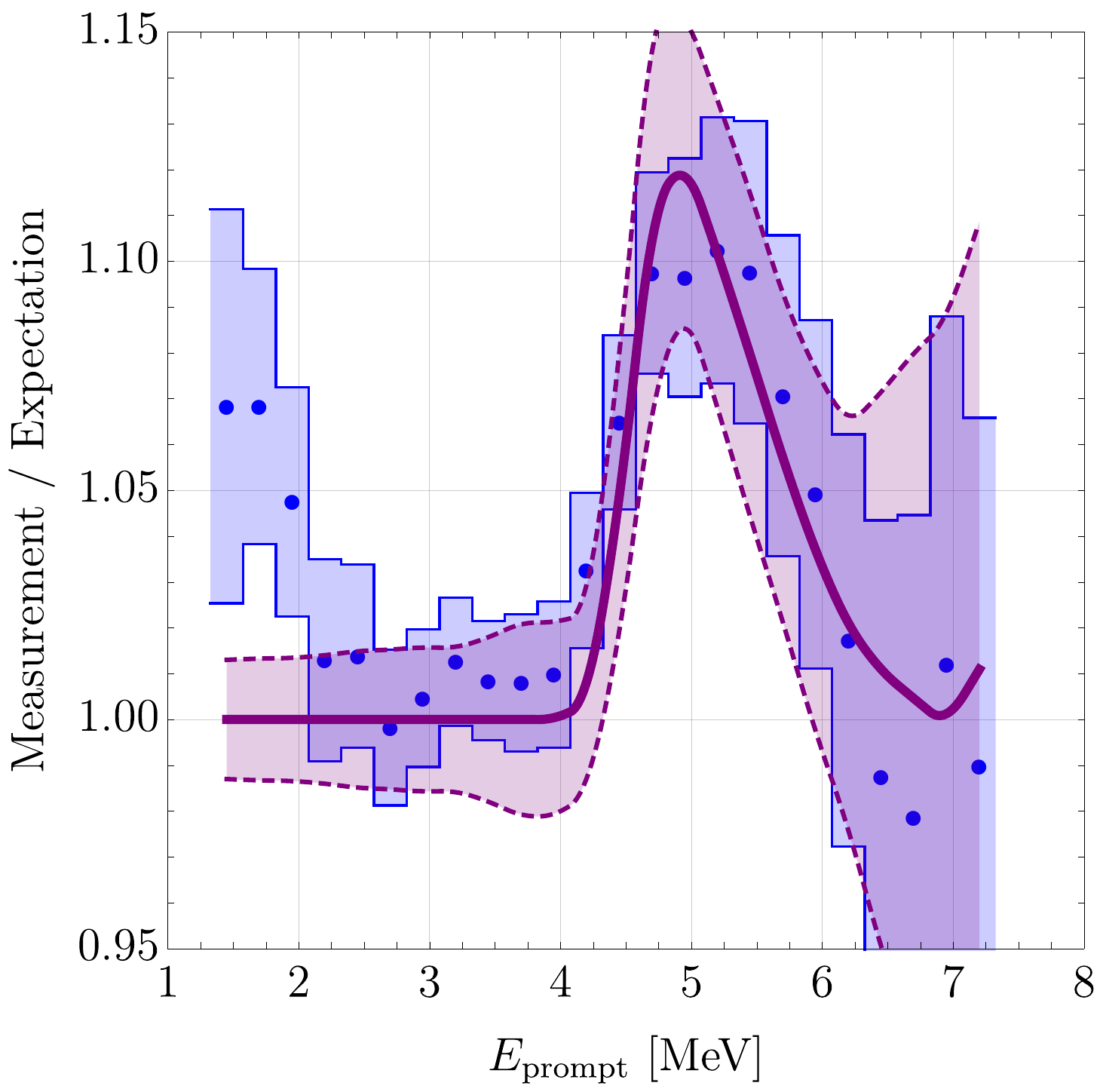}
\caption{The prompt energy spectrum corresponding to the best-fit point in \cref{eq:BFP}. Only transitions to the first excited state of $^{12}$C have been included. See text for details.}
\label{fig:results}
\end{figure}
\begin{figure}[!tbp]
\includegraphics[width=\linewidth]{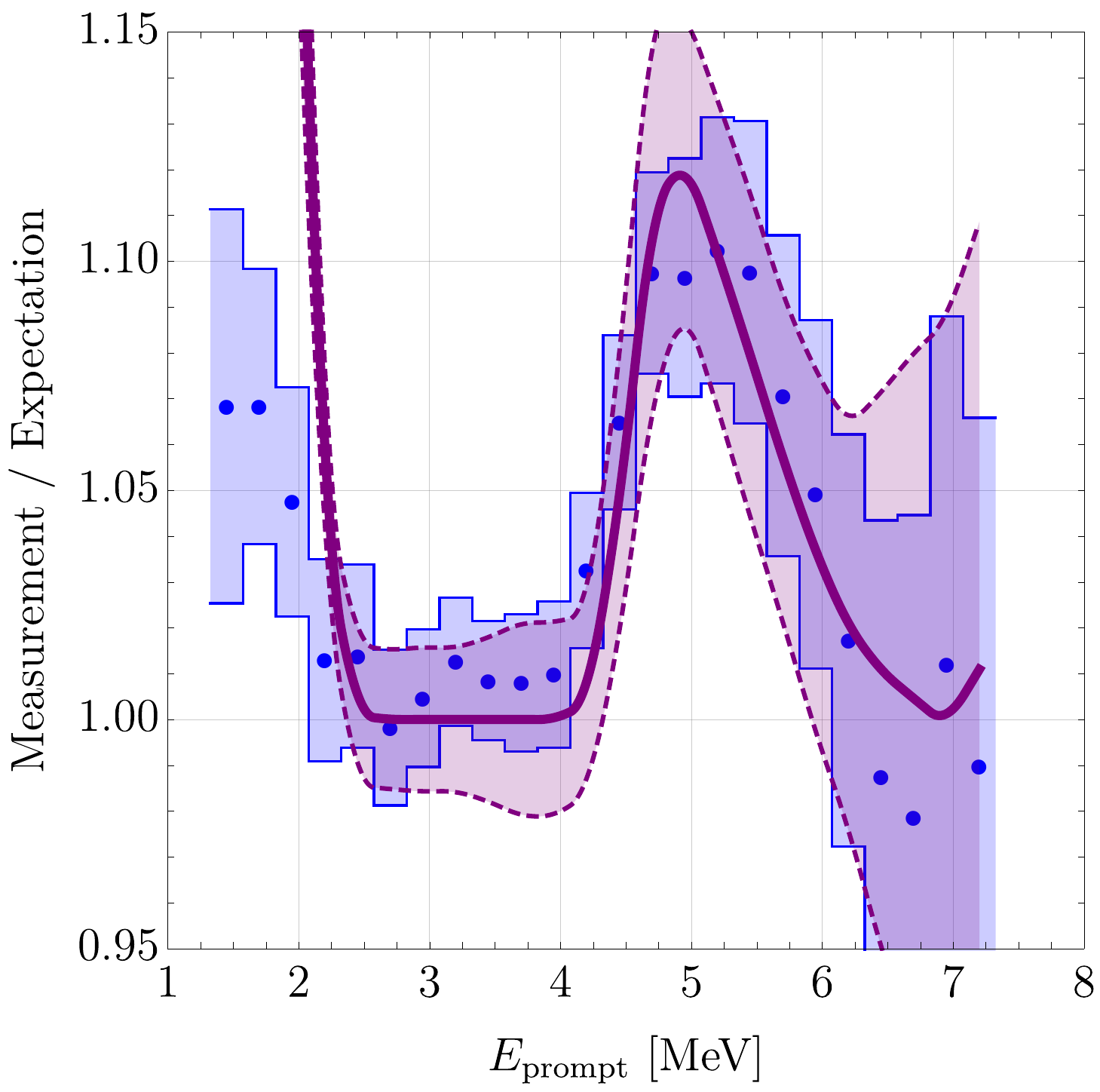}
\caption{The same as \cref{fig:results}, except that transitions to the ground state of $^{12}$C have been included. See text for details.}
\label{fig:results0}
\end{figure}

Considering only transitions to the excited state of $^{12}$C, the minimum value of the $\chi^2$, $\chi^2_{\rm min}$,  is 10.3 and occurs for
\begin{equation}
\label{eq:BFP}
\{ M_X, \, \log_{10} \zeta, \, I \} = \{ 1 \text{ keV}, \, -6.075, \, 13 \};
\end{equation}
for 21 degrees of freedom (d.o.f.), this is equivalent to $p = 0.97$.\footnote{This point is on the boundary of our scan, but we have verified that the minimum $\chi^2$ for a given value of $M_X$ in this range is not appreciably larger than the global minimum $\chi^2$.} The curve corresponding to this best-fit point is shown in \cref{fig:results}. The blue circles and blue bands correspond to the Daya Bay data and their associated uncertainties; the solid, purple line is calculated for the parameters in \cref{eq:BFP} and the dotted, purple lines show the uncertainties on the neutrino spectrum and the measurements of $A_0(E_{n})$, which have been added in quadrature. Marginalizing over the power-law index gives $\chi^2_{\rm min}$/d.o.f.~= 9.1/22 ($p = 0.99$), compared to $35.1/24$ ($p = 0.07$) in the absence of any NP.

When transitions to the ground state of $^{12}$C are included, the best-fit prompt energy spectrum presents a steep upturn at low energies; the fit cannot reproduce the 5 MeV bump without overproducing low-energy events by a factor of $\sim10$, as shown in \cref{fig:results0}. For the best-fit point in \cref{eq:BFP}, including transitions to the ground state increases $\chi^2_{\rm min}$/d.o.f. to $49.1/21$ ($p = 4.9 \times 10^{-4}$). We estimate that a $\lesssim \mathcal{O}(10\%)$ suppression of these transitions is sufficient to avoid problems. For a purely vector interaction, such a suppression is entirely {\it ad hoc}. However, if the new interaction were axial in nature, then it is likely to significantly suppress (or eliminate altogether) such transitions: the $^{13}$C ground state has $J^\pi = 1/2^{-}$, whereas the $^{12}$C ground state has $J^\pi = 0^{+}$ and the first excited state in $^{12}$C has $J^\pi = 2^{+}$. A full calculation using the $^{13}$C and $^{12}$C wave functions is beyond the scope of this work. 

We have also explored the effects of a threshold\footnote{The origin of such a threshold could stem for instance from an up-scattering process (which would require the presence of MeV-scale partner of a sterile neutrino).} $Q_{\text{th}}^2$ in the momentum exchange $-q^2=2 E_i E_f (1-\cos \theta_f)$ to determine if this can lead to the required suppression of the ground state transitions, which are generally characterized by smaller momentum exchanges. The effects of a threshold can be equivalently represented by an upper bound $\cos \theta_f < 1 - Q_{\rm th}^2/2E_i E_f$. 
The total Mott cross section becomes
\begin{align}
\sigma_{\rm Mott} = & \, \, \frac{\pi \alpha^2 Z^2}{E_i E_f} \times \left[ \log \left( \frac{4 E_i E_f + M_X^2}{M_X^2+Q_{\rm th}^2} \right) \right. \nonumber\\ 
& \left. - \frac{M_X^2 (4 E_i E_f-Q_{\rm th}^2)}{(4 E_i E_f + M_X^2)(M_X^2+Q_{\rm th}^2)} \right];
\end{align}
in contrast with \cref{eq:Mott}, in which no threshold is present.

\begin{table}[t]
\centering\begin{tabular}{|c|c|c|c|}\hline
Scenario & $\chi^2_{\rm min}$ & d.o.f. & $p$ \\ \hline \hline
No NP & 35.1 & 24 & 0.07 \\ \hline
NP, $Q_{\rm th}$ = 0 MeV & 16.8 & 22 & 0.77 \\ \hline
NP, $Q_{\rm th}$ = 5 MeV & 17.3 & 22 & 0.75 \\ \hline
NP, $Q_{\rm th}$ = 10 MeV & 17.1 & 22 & 0.76 \\ \hline
NP, $Q_{\rm th}$ = 15 MeV & 16.8 & 22 & 0.77 \\ \hline \hline
NP, No Ground State & 9.1 & 22 & 0.99 \\ \hline
\end{tabular}
\caption{The values of $\chi^2_{\rm min}$ for the cases in which NP is not present; where it is present, but there is no threshold $Q_{\rm th}$ in the momentum transfer; and in the cases where this threshold is 5, 10 or 15 MeV. We also show the number of degrees of freedom (``d.o.f.'')~in each fit, as well as the corresponding $p$-value. Transitions to the ground state of $^{12}$C are unsuppressed in these analyses. In the last row, we show the value of the $\chi^2_{\rm min}$ for the case in which transitions to the ground state are totally suppressed, for comparison.}
\label{tab:ChiSquareQth}
\end{table}

The results of this analysis are shown in Table~\ref{tab:ChiSquareQth}. The fits are obtained in a similar fashion as described for the scenario with no threshold, except that transitions to the ground state of $^{12}$C are unsuppressed. It is clear that nonzero $Q_{\rm th}$ has not appreciably lowered $\chi^2_{\rm min}$ relative to the case in which $Q_{\rm th} = 0$. Together with the increased challenge of constructing a model that would exhibit this kind of behavior (since stronger NP interactions would be required), this does not seem like a viable way to suppress ground-state transitions.

Taken at face value, including the ground-state transitions still seems to be a modest improvement over the SM prediction, even if ignoring them provides a better fit. However, most of the improvement to the fit in the former case comes from fitting the upturn at low energies in the Daya Bay data; the best-fit solution does little to reproduce the 5 MeV bump. We perform a similar analysis in which we omit the six lowest-energy Daya Bay data points to isolate the 5 MeV bump\footnote{We emphasize that we know of no reason why these data should be considered untrustworthy. We are simply restricting our analysis to focus on the 5 MeV bump instead of the low-energy upturn.}. With this restriction, the best-fit point is identical to that of \cref{eq:BFP} (independent of our previous analysis), and $\chi^2_{\rm min}$/d.o.f., having marginalized over the power-law index, is 1.4/16 ($1-p = 8.9 \times 10^{-7}$). The equivalent result in the absence of NP is 24.1/18 ($p = 0.09$). Importantly, this result is independent of whether or not ground-state transitions are included in the analysis. As such, the inclusion of ground-state transitions has no effect on the 5 MeV bump itself; they only cause the fit to struggle to reconcile the bump with the upturn in the low-energy data.

We briefly consider the NEOS experiment~\cite{Ko:2016owz} which sees a similar bump at 5\,MeV in their prompt energy spectrum. Including only transitions to the first excited state of $^{12}$C and marginalizing over the power-law index, we find $\chi^2_{\rm min}$/d.o.f.~$=14.2/58$ ($1-p=6.2 \times 10^{-10}$) at the best-fit point compared to $66.7/60$ ($p=0.26$) without NP. Moreover, the results of Daya Bay and NEOS are consistent with one another; analyzing both simultaneously gives $\chi^2_{\rm min}$/d.o.f.~$ = 101.8/84$ ($p=0.09$) in the absence of NP, while $\chi^2_{\rm min}$/d.o.f.~$ = 24.0/82$ ($1-p = 4.7\times10^{-11}$) is obtained with transitions only to the first excited state of $^{12}$C. We demonstrate the consistency between these two data sets quantitatively using a parameter goodness of fit test \cite{Maltoni:2003cu}. We calculate $\chi^2_{\rm PG} \equiv \chi^2_{\rm min, DB + NEOS} - \chi^2_{\rm min, DB} - \chi^2_{\rm min,NEOS} = 0.66$, which follows a $\chi^2$ distribution with the number of d.o.f.~given by the number of shared parameters in the fits of Daya Bay (DB) and NEOS, {\it i.e.}, two. The corresponding $p$-value is $0.72$, indicating a reasonable level of compatibility between these data sets.

\section{Constructing models for \boldmath{$^{13}$C$(\overline{\nu}, \overline{\nu}^\prime$\emph{\lowercase{n}}$)^{12}$C$^{(*)}$}}
\label{sec:models}

In searching for a viable model for $^{13}$C$(\overline{\nu}, \overline{\nu}^\prime n)^{12}$C$^*$, we first focus on the nuclear physics requirements. In order to produce a neutron and $^{12}$C$^*$, $^{13}$C nuclei need to reach an excited state in scattering with neutrinos. The first two excited states of $^{13}$C that could decay into $^{12}$C$^*$ have either opposite parity or different spin with respect to the ground state. This precludes a scalar from being the mediator of this interaction. Nevertheless, selection rules allow for the exchange of a boson, namely a spin-2 particle in such a process. The most notable example of a spin-2 state is the graviton. However, due to the nonrenormalizability of quantum gravity, this option is not appealing. The same holds for general spin-2 particles. The simplest option is thus the exchange of a new vector particle, which we have already hinted at in \cref{sec:xsec+flux} when obtaining the $^{13}$C$(\overline{\nu}, \overline{\nu}^\prime n)^{12}$C$^{(*)}$ scattering cross section in \cref{eq:Mott}. Admittedly, we are biased towards a pure vector interaction because, as already stated, estimating the scattering cross section requires use of available data from experiments --- to wit, we have used the cross section data for $^{13}$C$(e,e^\prime n)^{12}$C$^{(*)}$, which is mediated by the SM photon. 

We turn now to several concrete realizations of vector interactions.

\subsection{Vector interaction}
\label{subsec:vector}

We start by employing a generic dark photon coupled to the SM photon via kinetic mixing $\epsilon$ \cite{Nelson:2011sf},
\begin{align}
\mathcal{L} & =-\frac{1}{4} F^{\mu\nu} F_{\mu\nu}-\frac{1}{4} F'^{\mu\nu} F'_{\mu\nu}-\frac{\epsilon}{2} F^{\mu\nu}F'_{\mu\nu} +\frac{1}{2} M_X^2 X^2 \nn \\
& - e \sum_\psi Q_\psi A_\mu \overline{\psi} \gamma^\mu \psi,
\end{align} 
where $\psi$ is a SM fermion with charge $Q_\psi$ in units of $e$, $F^{\mu\nu}$ and $F'^{\mu\nu}$ are the photon $(A)$ and dark photon $(X)$ field strength tensors, respectively. Since SM neutrinos are not charged under electromagnetism, they do not develop a coupling to the dark photon. This minimal setup cannot explain the reactor bump.

We move on to models in which SM neutrinos are charged under an
extra Abelian gauge group so that they couple to $X$ bosons. An
attractive option is $U(1)_{B-L}$, which is anomaly free once
right-handed neutrinos are introduced
\cite{Buchmuller:1991ce,Khalil:2010iu,Sahu:2004sb}. This model yields
$\zeta^2 \sim \left (g_{B-L}\right)^2 \epsilon^2$ or $\left(
g_{B-L} \right)^4$, depending on the relative sizes of the
kinetic mixing and the $B-L$ coupling constant. Unfortunately, this
model faces stiff constraints, as every other SM fermion is also
charged under this group; see
Refs.~\cite{Harnik:2012ni,Cerdeno:2016sfi} for details.

We are then led to an alternative Abelian symmetry, which we call
$U(1)_X$. We introduce a fourth species of neutrino ($\nu_s$) with
nonzero charge under $U(1)_X$, and allow for nucleons to also be
charged. The relevant part of the interaction Lagrangian is
\begin{align}
\mathcal{L} = g_X X_\mu ( Y_\nu \,\overline{\nu}_s \gamma^\mu \nu_s + \overline{p} \gamma^\mu p + Y_n \, \overline{n} \gamma^\mu n),
\end{align}
where $g_X$ is the $U(1)_X$ coupling constant, and $Y_\nu$, $Y_n$ are
the new neutrino and neutron charge, respectively, in units of $g_X$. When $Y_n =
+1$, $U(1)_X$ becomes gauged baryon number, $U(1)_B$, considered in
Ref.~\cite{Pospelov:2011ha}. The phenomenology of an eV-scale neutrino
charged under $U(1)_B$ was studied in Ref.~\cite{Kopp:2014fha}. We
identify
\begin{equation}
\label{eq:u1b}
\zeta^2 = \sin^2 \left(\frac{\Delta m_{41}^2 L}{4 E_\nu}\right) \sin^2 2\theta_{ee} \left( \frac{6+7Y_n}{6} \right)^2 Y_\nu^2 \left(\frac{g_X}{e} \right)^4,
\end{equation}
where $\Delta m_{41}^2$ is the new mass-squared splitting and $\theta_{ee}$ is the new mixing angle. For consistency with short-baseline oscillation measurements \cite{Kopp:2013vaa,Gariazzo:2017fdh,Dentler:2017tkw,Gariazzo:2018mwd,Dentler:2018sju}, we take $\Delta m^2_{41} = 0.4$ eV$^2$ and $\sin^2 2\theta_{ee} = 0.04$ as our benchmark values for deriving limits and best-fits, unless otherwise stated. Given the baseline and characteristic neutrino energy, $\Delta m_{41}^2$ is sufficiently large so that its oscillations average out at Daya Bay, allowing us to replace $\sin^2 \left(\frac{\Delta m_{41}^2 L}{4 E_\nu}\right) \to \frac{1}{2}$ in \cref{eq:u1b}.

\begin{figure}[!tbp]
\includegraphics[width=\linewidth]{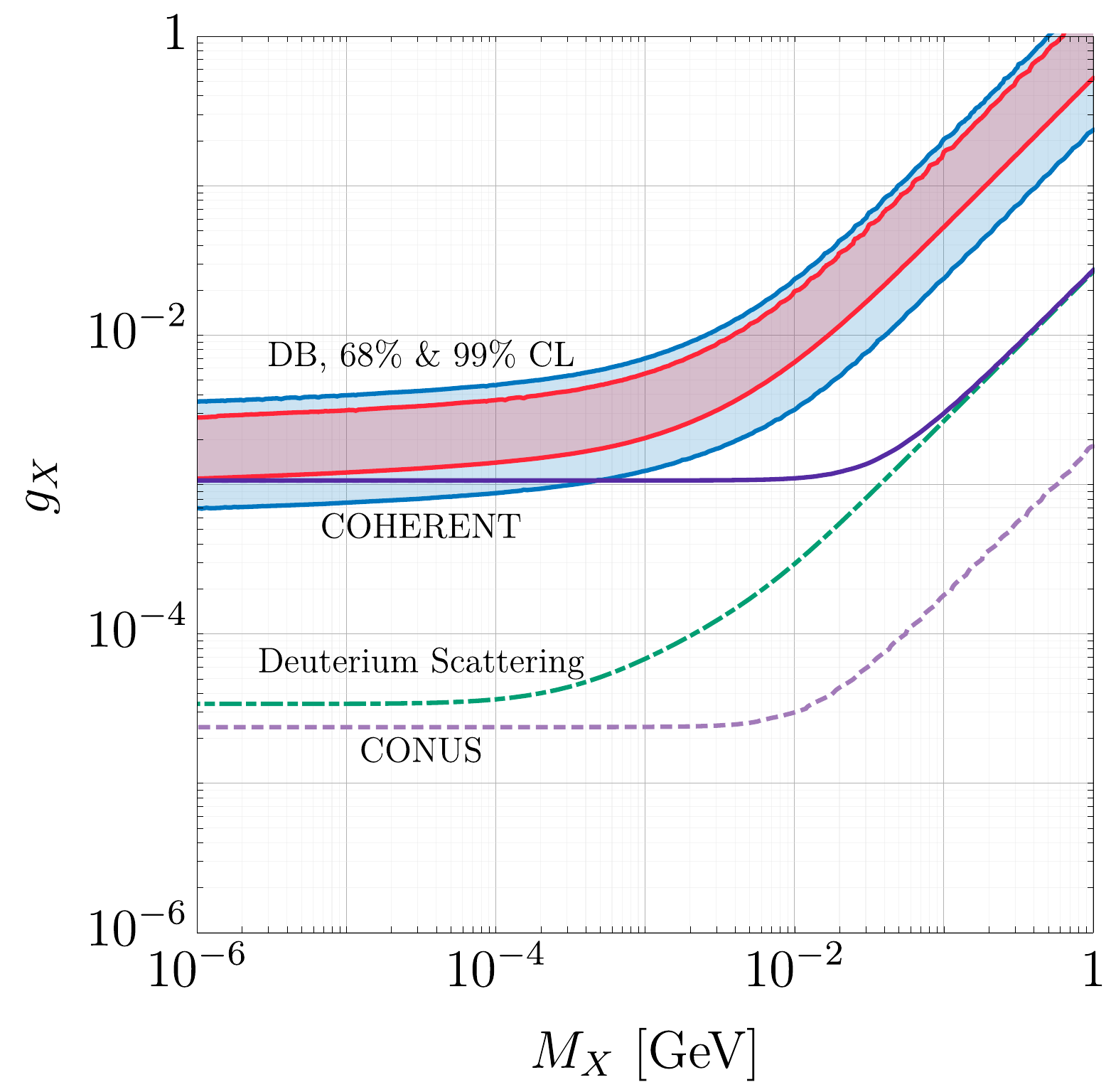}
\caption{Limits on the $U(1)_X$ model with $Y_\nu$ = +10, $Y_n = -0.65$. The solid purple (dashed violet) line is the exclusion limit (sensitivity reach) of COHERENT (CONUS), whereas the green, dot-dashed line is the approximate exclusion from neutrino-induced deuterium disintegration. The red (blue) band is the 68\% (99\%) confidence level preferred by the Daya Bay data; we have marginalized over the power-law index, and are suppressing transitions to the ground state of $^{12}$C.}
\label{fig:bounds}
\end{figure}
\begin{figure}[!tbp]
\includegraphics[width=\linewidth]{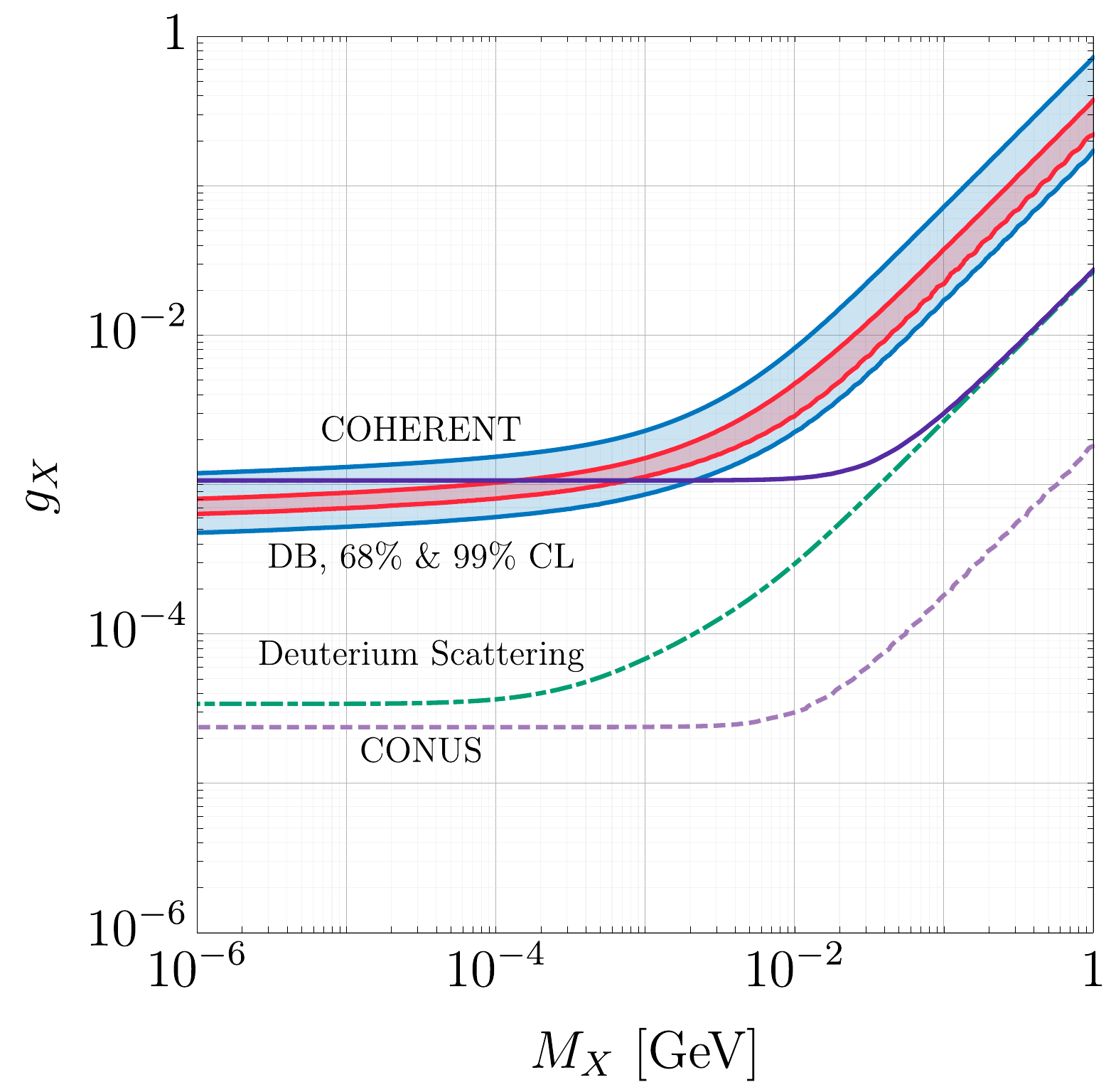}
\caption{The same as \cref{fig:bounds}, except we have included transitions to the ground state of $^{12}$C in our fit. While more of the region preferred by Daya Bay lies beyond the exclusion reach of COHERENT, the overall quality of the fit is poorer than in fig.~\ref{fig:bounds} ($\chi_{\rm min}^2$ = 16.8 vs. 9.1); see text for details.}
\label{fig:bounds0}
\end{figure}

The red (blue) band in \cref{fig:bounds} represents the 68\% (99\%) confidence level (CL) from Daya Bay data derived by the analysis described in the previous section, after marginalizing over the power-law index $I$.  In our analysis, we have chosen $Y_n=-0.65$ and $Y_\nu=10$. Constraints from pion decay ($\pi^0 \to \gamma \, + $ invisible) \cite{Atiya:1992sm,Batell:2014yra,deNiverville:2015mwa} imply $g_X \gtrsim 10^{-2}$ for $M_X < m_\pi$ and depend on neither $Y_n$ nor $Y_\nu$; as long as the perturbativity condition $Y_\nu \, g_X < 4\pi$ is maintained, the new neutrino charge can be arbitrarily large. Choosing $Y_\nu=10$ allows us to circumvent these constraints. Moreover, there exist strong bounds on this scenario from $^{208}$Pb-n scattering \cite{Barbieri:1975xy,Leeb:1992qf}. These bounds, however, can be evaded by judiciously choosing the neutron charge to minimize the charge of $^{208}$Pb under this interaction; the choice $Y_n=-0.65$ allows us to do precisely this. We have also considered bounds from kaon decay \cite{Ecker:1987qi,Ecker:1987hd,DAmbrosio:1998gur,Pospelov:2008zw} and from fifth-force searches \cite{Harnik:2012ni,Cerdeno:2016sfi}; these do not provide particularly stringent additional constraints. 

The most relevant bounds on this model derive from neutrino scattering experiments. We show three such constraints in \cref{fig:bounds}: bounds from COHERENT \cite{Akimov:2017ade} (purple), CONUS \cite{CONUStalk} (dashed violet) and deuterium scattering \cite{Reines:1980pc} (dot-dashed green). While COHERENT leaves a portion of the preferred Daya Bay region unconstrained, the measurement of the deuterium disintegration fully excludes this model. In the near future, this exclusion is expected to be even stronger, once the results of the CONUS experiment, which is probing coherent neutrino-nucleus scattering at a reactor source, are released.

In \cref{fig:bounds0}, we show the results of a similar analysis to \cref{fig:bounds}, except transitions to the ground state of $^{12}$C are included without suppression. While more of the parameter space preferred by Daya Bay lies beyond the exclusion reach of COHERENT, we do not regard this as a genuine improvement over \cref{fig:bounds} for two reasons. (1) The overall quality of the fit is poorer. The minimum chi-squared, $\chi^2_{\rm min}$, is 16.8 here, whereas this value is 9.1 when transitions to the ground state are removed. This is because the fit prioritizes the low-energy upturn over the bump at 5 MeV, as mentioned above. (2) The power-law index $I$ preferred by the fit is smaller ($I \sim 7$ vs. $I\sim13$), indicating a preference for a harder antineutrino spectrum at high energies. We have not formally constrained this high-energy component in our fit, but preliminary data \cite{NRThesis} suggest that this is inconsistent with Daya Bay observations. Therefore, we will continue to ignore transitions to the ground state of $^{12}$C for the rest of this work.

In what follows, we briefly discuss our treatment of coherent elastic neutrino-nucleus scattering (CE$\nu$NS) and neutrino-induced deuteron disintegration.

\subsubsection*{COHERENT and CONUS}

The differential cross section for CE$\nu$NS involving $\nu_\alpha$ (or $\overline{\nu}_\alpha$) is
\begin{equation}
\label{eq:CEnNS}
\frac{d\sigma_\alpha}{dE_r} = \frac{G_F^2}{\pi} |Q_\alpha|^2 F^2(q^2) M \left(1 - \frac{M E_r}{2E_\nu^2} \right),
\end{equation}
where $E_\nu$ is the neutrino energy, $G_F$ is Fermi's constant, $M$ is the mass of target nucleus with $N$ neutrons and $Z$ protons, $q^2 \approx -2 M E_r$ is the momentum transfer that gives rise to nuclear recoil energy $E_r$, $F^2(q^2)$ is the Helm form factor \cite{Klein:1999gv} and $Q_\alpha$ is the species-dependent effective charge. In the Standard Model, $Q_\alpha$ is given by
\begin{equation}
Q_\alpha = Z g_p^V + N g_n^V,
\end{equation}
where $g_p^V \, = \frac{1}{2} - 2 \sin^2 \theta_W$ and $g_n^V \, = -\frac{1}{2}$ are the proton and neutron weak vector couplings, respectively, and $\theta_W$ is the Weinberg angle. We ignore contributions from the axial part of the weak interactions. 

The presence of a light new vector particle induces additional contributions to the effective charge that interfere with the SM. These contributions to $Q_\alpha$ are weighted by the relevant oscillation amplitudes. Namely,
\begin{align}
\label{eq:defineQ}
Q_\alpha = & \, \, \mathcal{A}_{\alpha \alpha} (Z g_p^V + N g_n^V) \nonumber \\
& + \mathcal{A}_{\alpha s} \left( \frac{-\sqrt{2} g_X^2 Y_{\nu,\overline{\nu}} (Z+Y_n N)}{2G_F (M_X^2 + 2 M E_r)} \right),
\end{align}
where $Y_{\nu,\overline{\nu}}$ is either the neutrino or antineutrino charge, as appropriate; $\mathcal{A}_{\alpha \alpha}$ is the amplitude for $\nu_\alpha$ disappearance; and $\mathcal{A}_{\alpha s}$ is the amplitude for $\nu_s$ appearance in a $\nu_\alpha$ beam. In the limit in which oscillations among the SM neutrinos are irrelevant, the oscillation amplitudes are
\begin{align}
\label{eq:amps}
\mathcal{A}_{\alpha \alpha} & = 1 - 2 i \exp\left({\frac{-i \Delta m^2_{41} L }{4 E_{\nu}}}\right) \sin^2 \theta_{\alpha \alpha} \sin\left( {\frac{\Delta m^2_{41} L }{4 E_{\nu}}}\right) \\
\mathcal{A}_{\alpha s} & = 2 i \exp\left({\frac{-i \Delta m^2_{41} L }{4 E_{\nu}}}\right) \cos \theta_{\alpha \alpha} \sin \theta_{\alpha \alpha} \sin\left( {\frac{\Delta m^2_{41} L }{4 E_{\nu}}}\right),
\label{eq:amps2}
\end{align}
where $\sin^2 \theta_{\alpha \alpha} \equiv |U_{\alpha 4}|^2$. Squaring this yields
\begin{align}
|Q_\alpha|^2 = & \, \, P_{\alpha \alpha} \times \left(Z g_p^V + N g_n^V \right)^2 \nonumber \\
& + P_{\alpha s} \times \left( \frac{-\sqrt{2} g_X^2 Y_{\nu,\overline{\nu}} (Z+Y_n N)}{2G_F (M_X^2 + 2 M E_r)} \right)^2 \nonumber \\
& + 2 \sin 2\theta_{\alpha \alpha} \cos 2\theta_{\alpha \alpha} \sin^2 \left( \frac{\Delta m_{41}^2 L}{4 E_{\nu}} \right) \times \nonumber \\
& \left(Z g_p^V + N g_n^V \right) \left( \frac{-\sqrt{2} g_X^2 Y_{\nu,\overline{\nu}} (Z+Y_n N)}{2G_F (M_X^2 + 2 M E_r)} \right),
\end{align}
where the oscillation probabilities are given by
\begin{equation}
P_{\alpha s} = 1 - P_{\alpha \alpha} = \sin^2 2\theta_{\alpha \alpha} \sin^2 \left( \frac{\Delta m_{41}^2 L}{4 E_{\nu}} \right).
\end{equation}
We emphasize that the only nonzero new mixing parameter that we consider is $\sin^2 2\theta_{ee}$, {\it i.e.}, muon neutrinos do not mix with the fourth flavor.\\

We wish to determine the exclusion reach of COHERENT and the sensitivity of CONUS. We simulate COHERENT following the procedure described in Refs.~\cite{Liao:2017uzy,Farzan:2018gtr}; for CONUS, we follow Ref.~\cite{Farzan:2018gtr}. For most values of $Y_n$, COHERENT excludes most of Daya Bay's preferred parameter space. However, choosing the neutron charge to minimize the $U(1)_X$ charges of the constituent nuclei allows for the bound to be loosened by a factor of $\sim 10$. Since the relevant isotopes for COHERENT are $^{133}$Cs and $^{127}$I, $Y_n = -0.65$ -- which we had previously chosen to evade $^{208}$Pb-n scattering constraints -- provides modest relief (see \cref{eq:u1b}).

We show the 99\% CL exclusion limit for COHERENT in purple in \cref{fig:bounds}. If we had selected the best-fit point from Ref.~\cite{Dentler:2018sju}, $( \Delta m_{41}^2, \, \sin^2 2\theta_{ee}) \sim (1.3 \text{ eV}^2, \, 0.04)$, then COHERENT would have ruled out the entire region preferred by Daya Bay. Given COHERENT's characteristic neutrino energy ($\sim \mathcal{O}(30)$ MeV) and baseline (19.3 m), the event rate approximately scales as $g_X^4 \sin^2 2\theta_{ee} (\Delta m_{41}^2)^2$ for $\Delta m_{41}^2 \lesssim 0.5$ eV$^2$. The COHERENT constraint is weakened as $\Delta m^2_{41}$ is decreased, but our willingness to do so in order to avoid this constraint is dictated by reactor data \cite{Dentler:2018sju}. Our benchmark parameters -- $( \Delta m_{41}^2, \, \sin^2 2\theta_{ee}) \sim (0.4 \text{ eV}^2, \, 0.04)$ -- are allowed at 95\% CL by Ref.~\cite{Dentler:2018sju} and permit us to marginally evade this constraint. 

While the sensitivity of COHERENT is slightly diminished for $Y_n = -0.65$, the CONUS detector is made of natural germanium, for which the proton-to-neutron ratio is $\sim 0.8$. Consequently, each of the five naturally-occurring germanium isotopes has nonzero charge under $U(1)_X$, so it is not possible to simultaneously provide relief from both Pb-n scattering and COHERENT constraints while also limiting the sensitivity of CONUS. The 99\% CL sensitivity of CONUS is shown in dashed violet in \cref{fig:bounds}.

\subsubsection*{Neutrino-induced deuterium disintegration}

With the COHERENT limit not fully excluding the preferred Daya Bay region, and with CONUS not yet having released any results, only the bound from the neutrino-induced deuterium disintegration currently rules out the exchange of a vector boson. We briefly discuss how this limit has been calculated.

We have calculated the contributions of our NP model to the cross section for inelastic neutrino-deuterium scattering at reactors, following the calculations of Refs.~\cite{Butler:1999sv,Butler:2000zp}. In \cref{app:deuterium}, we outline the computation of the cross section in the SM and demonstrate how the effects from NP can be included.

To obtain the limit, we first calculate the flux-weighted cross section at 11.2 m from a nuclear reactor, as in Ref.~\cite{Reines:1980pc}, for our benchmark fourth neutrino parameters using the HM fluxes for a range of $g_X$ and $M_X$. The results are then compared against the same for the purely SM case. Ref.~\cite{Reines:1980pc} reports a measurement of the $\overline{\nu}$ scattering rate consistent with the SM value at the $\sim10\%$ level; this allows us to exclude parts of the $\{g_X, \, M_X\}$ parameter space. In \cref{fig:bounds}, we show in light green the contour along which $\tilde{\sigma}^{\rm SM+NP} = 2\tilde{\sigma}^{\rm SM}$ in the $\{g_X, \, M_X\}$ plane, where $\tilde{\sigma}^{\rm SM}$ ($\tilde{\sigma}^{\rm SM+NP}$) is the SM (SM+NP) flux-weighted cross section; this is intended to be a conservative limit, given the complexity of the underlying measurement. This result clearly excludes the vector-boson-exchange explanation of the reactor antineutrino excess.

\subsection{Axial interaction}
\label{subsec:axial_vector}

Next, we consider a spin-1 boson $X$ with purely axial couplings to the fourth neutrinos and to nucleons,
\begin{equation}
\mathcal{L} \supset g_X X_\mu \left( Y_\nu \overline{\nu}_s \gamma^\mu \gamma_5 \nu_s + \overline{p} \gamma^\mu \gamma_5 p + Y_n \overline{n} \gamma^\mu \gamma_5 n \right)\,.
\end{equation}
Integrating out $X$ leads to the effective Lagrangian of the form
\begin{equation}
\label{eq:Laxial}
\mathcal{L} \supset \frac{-g_X^2 Y_\nu}{q^2-M_X^2} \left(\overline{\nu}_s \gamma^\mu P_L \nu_s\right) \left( \overline{p} \gamma_\mu \gamma_5 p +Y_n \overline{n} \gamma_\mu \gamma_5 n \right),
\end{equation}
where we have selected the left-handed part of the neutrino current because any neutrino present in an experiment will arise from a neutrino (antineutrino) produced in weak-interaction processes and will thus be initially left-handed (right-handed), to leading order.

Naively, an axial-vector interaction may be able to relax constraints from COHERENT and CONUS; for similar vector and axial couplings to nucleons, the axial contribution to the cross section is roughly a factor of $1/A$ smaller than its vector counterpart \cite{Barranco:2005yy,Scholberg:2018vwg}. Hence, if the NP vector couplings were to vanish, then the axial interaction may be significantly less constrained than a comparably-sized vector interaction. We have performed such an analysis and have found a factor of few relaxation of the $g_X$ limit. This qualitatively does not alter the picture given in \cref{subsec:vector}. It is interesting to note that it is possible to make either the COHERENT (CONUS) bound vanish, at leading order, by demanding the proton (neutron) to be uncharged under this new axial interaction. 

Despite the more positive outlook at CE$\nu$NS experiments compared to the pure vector interaction, the bound from deuterium disintegration turns out to be stronger than its vector-interaction counterpart. We do not review here the full derivation, but point the interested reader to \cref{app:deuterium}, where the treatment for the vector interaction is outlined. We emphasize that only the coefficients representing the axial-vector isoscalar and isovector couplings $(C_A^{(0)}, C_A^{(1)})$ get modified in this case. As mentioned previously, a full calculation of the $^{13}$C$(\overline{\nu}, \overline{\nu}^\prime n)^{12}$C$^{(*)}$ cross section for an axial-vector mediator is well beyond the scope of this work. However, it is likely that neutrino-induced deuterium disintegration would place a cripplingly strong bound on such a scenario.

\subsection{Magnetic interaction}
\label{subsec:magnetic}
 
Lastly, we consider a magnetic-magnetic interaction between the new neutrinos and nuclei. We introduce an effective interaction of the form
\begin{equation}
\label{eq:magnetic}
\mathcal{L} \supset \sum_{N = p, \, n} \frac{\mu_\nu \mu_N}{q^2-M_X^2} \left( \overline{\nu}_s \sigma^{\mu \nu} (i q_\nu) \nu_s \right) \left( \overline{N} \sigma_{\mu \alpha} (i q^\alpha) N \right),
\end{equation}
where $\mu_\nu$ and $\mu_N$ are the neutrino and nucleon magnetic dipole moments, respectively. We parameterize the latter as $\mu_N = g_X \mu^{\rm SM}_N$, where $\mu^{\rm SM}_N$ is the SM magnetic dipole moment with the fundamental charge $e$ factored out ($\mu^{\rm SM}_p = 2.79/2m_N$ and $\mu^{\rm SM}_n = -1.91/2m_N$). The neutrino magnetic moment is defined to be $\mu_\nu \equiv g_X Y_\nu/M_X$.

To estimate the constraint from Daya Bay, we follow the example of Ref.~\cite{Grifols:2004yn}, wherein the $\overline{\nu}d \to \overline{\nu}np$ cross section is calculated for nonzero neutrino magnetic moments using the $\gamma d \to np$ cross section to sidestep complications arising from nuclear structure. Of course, the $^{13}$C$(X, n)^{12}$C$^{(*)}$ cross section is not measured; we use the $^{13}$C$(\gamma, n)^{12}$C$^{(*)}$ cross section instead \cite{Koning:2012zqy} modified by a factor of $\frac{2}{3}(g_X/e)^2$, to account for (1) the difference in coupling strengths, and (2) the difference in the number of allowed spin states of our massive vector boson. This forces us into assuming that the longitudinal mode of our massive boson would not contribute appreciably to $^{13}$C$(X, n)^{12}$C$^{(*)}$. The experimentally determined total $^{13}$C$(\gamma, n)^{12}$C$^{(*)}$ cross section does not reveal how much of this cross section comes from M1 scattering and how much from E1 scattering. We assume that M1 scattering dominates; if this is untrue, then the Daya Bay data will prefer larger values of $g_X$ than this optimistic analysis.

With these assumptions and by following Ref.~\cite{Grifols:2004yn}, we obtain the following differential cross section:
\begin{widetext}
\begin{equation}
\label{eq:DBmagnetic}
\frac{d\sigma}{d E^\prime} = \frac{\mu_\nu^2 g_X^2 (E-E^\prime) E^\prime \times \sigma_\gamma(E_\gamma = E - E^\prime)}{4 \pi^3 \alpha E (M_X^2+2(E-E^\prime)^2) (M_X^2+4 E E^\prime)} \times \Big[ 4 E E^\prime (M_X^2+2 E E^\prime) - M_X^2 (M_X^2+4 E E^\prime) \log\left( \frac{M_X^2+4 E E^\prime}{M_X^2} \right) \Big]\,,
\end{equation}
\end{widetext}
where $E$ ($E^\prime)$ is the initial (final) antineutrino energy and $\sigma_\gamma$ is the $^{13}$C$(\gamma, n)^{12}$C$^{(*)}$ cross section. Note that the final neutrino energy satisfies $E^\prime = E - S_n -E_{\rm ex} - E_{\rm nuc}$, where $S_n$ is the neutron removal energy of $^{13}$C, $E_{\rm ex}$ is the excitation energy of the final $^{12}$C (either 0 or $\sim$ 4.4 MeV), and $E_{\rm nuc}$ is the kinetic energy carried off by the $^{12}$C-neutron system. We assume that the neutron carries off most of this energy and, therefore, that \cref{eq:DBmagnetic} is also the neutron recoil spectrum. 

\begin{figure}[!t]
\includegraphics[width=\linewidth]{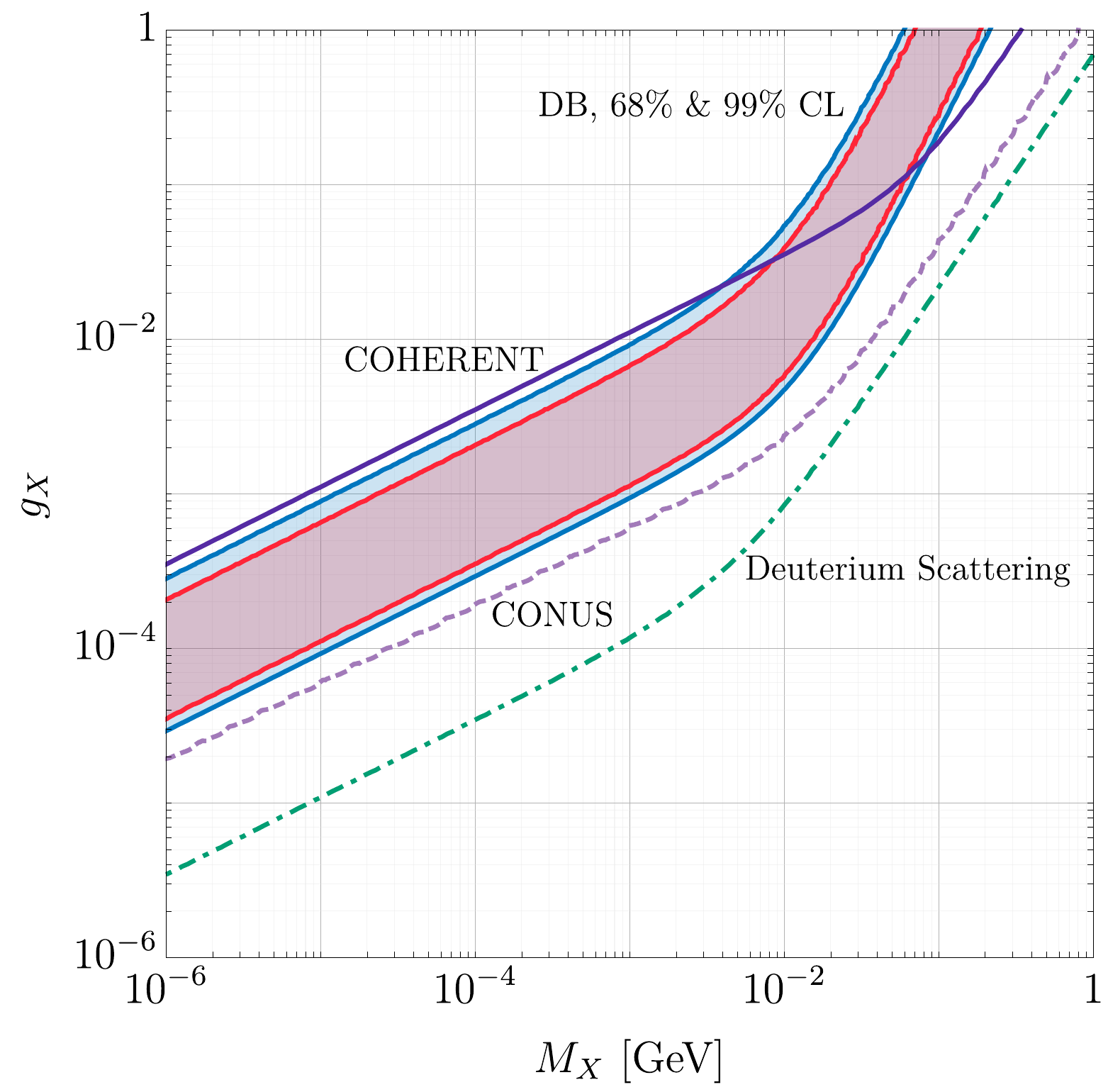}
\caption{Bounds on the magnetic-interaction scenario from COHERENT (solid purple), CONUS (dashed violet) and deuterium disintegration (dot-dashed green), as described in the text. We continue to take $\Delta m_{41}^2 = 0.4$ eV$^2$, $\sin^2 2\theta = 0.04$ and $Y_\nu = +10$. Here, we include only transitions to the first excited state of $^{12}$C in our analysis of Daya Bay.}
\label{fig:magnetic}
\end{figure}

The red (blue) band in \cref{fig:magnetic} is the 68\% (99\%) confidence level preferred by the Daya Bay data. We include only transitions to the first excited state of $^{12}$C, and find that $\chi^2_{\rm min}$/d.o.f.~= 11.7/22 ($p=0.96$). Including transitions to the ground state of $^{12}$C makes the parameter space preferred by Daya Bay slightly narrower and increases $\chi^2_{\rm min}$/d.o.f.~$\to$ 16.6/22 ($p=0.78$); the fit still suffers from fitting the low-energy upturn instead of the 5 MeV bump. The no-ground-state fit is slightly poorer than its vector-interaction analog -- recall that we had found $\chi^2_{\rm min}$/d.o.f.~= 9.1/22 -- but it remains to determine the extent to which CE$\nu$NS and deuterium disintegration experiments can exclude such scenario, as in \cref{subsec:vector,subsec:axial_vector}. 

For the CE$\nu$NS cross section, we extend the formalism for nonrelativistic dark matter-nucleon effective field theory presented in Refs.~\cite{Fitzpatrick:2012ix,Anand:2013yka} to account for the fact that neutrinos are ultrarelativistic; a detailed calculation will be presented in Ref.~\cite{upcoming}. We have verified that we recover the usual CE$\nu$NS cross section from this framework in the absence of NP. The result is similar to the one presented in \cref{subsec:vector} for a vector interaction. In \cref{fig:magnetic}, we see that COHERENT (purple) does little to challenge the parameter space preferred by Daya Bay. We have shown the result for our benchmark fourth neutrino parameters, in the interest of consistency, but have determined that even for the best-fit point of Ref.~\cite{Dentler:2018sju}, having a larger $\Delta m_{41}^2$ ( = 1.3 eV$^2$), COHERENT provides only a modest exclusion of the high-mass ($M_X \gtrsim 10$ MeV) portion of the Daya Bay preferred region. However, CONUS (dashed violet) will have the power to fully exclude this model.

For deuterium disintegration, we repurpose the results of Ref.~\cite{Akhmedov:1991rt} to determine the cross section. We use a modified version of the M1 matrix element given in their eq.~(11), to account for differences in normalization convention and the effects of a nonzero $M_X$:
\begin{align}
|M|^2 = & \, \, |\mathcal{M_{\rm nuc}}|^2 g_X^2 (\mu^{\rm SM}_p-\mu^{\rm SM}_n)^2 \mu_\nu^2 \times \\
& E E_f \left( \frac{1+\mathbf{n_1} \cdot \mathbf{n_2}}{3} \right) \times (2 m_N)^3 \left(\frac{q^2}{q^2-M_X^2}\right)^2, \nonumber
\end{align}
where $\mathbf{n_1}$ ($\mathbf{n_2}$) is the unit vector along the direction of the incident (outgoing) neutrino, which respectively have energies $E$ and $E_f$. Ref.~\cite{Akhmedov:1991rt} only includes the isovector contributions to the scattering cross section because these would ultimately dominate. Additionally, $|\mathcal{M_{\rm nuc}}|^2$ incorporates the effects of the structure of deuterium; it is given by Eq.~(12) of Ref.~\cite{Akhmedov:1991rt} and depends on $E_k$, the kinetic energy of the relative motion of the final-state proton and neutron. 

The differential cross section reads
\begin{equation}
\label{eq:magneticD}
\frac{d\sigma}{dE_k} = \frac{|\mathcal{M}|^2 g_X^4 Y_\nu^2 \mu_{\rm SM}^2 (E-B-E_k)^2 \sqrt{E_k}}{384 \pi^3 \sqrt{m_N} M_X^2} f(z)\,,
\end{equation}
where $\mu_{\rm SM} \equiv 2m_N(\mu^{\rm SM}_p-\mu^{\rm SM}_n)$ is the isovector nucleon magnetic moment, $m_N$ is the nucleon mass and $B$ is deuteron binding energy. The function $f(z)$ reads
\begin{align}
 f(z) \equiv \int^1_{-1} dx \frac{(1+x)(1-x)^2}{(1+z-x)^2} = \\
 2+6z-2z(4+3z) \coth^{-1}(1+z)\,,
\end{align}
where we define $z \equiv M_X^2/(2E E_f)$.

From \cref{eq:magneticD}, it is straightforward to calculate the flux-weighted cross section and obtain the limit in a similar fashion as described at the end of \cref{subsec:vector}. From \cref{fig:magnetic}, we observe that the deuterium disintegration limit (dot-dashed green) again rules out the entirety of the parameter space preferred by Daya Bay. Hence, this model is strongly disfavored.

\section{Summary and Conclusions}

In this work, we have studied whether the 5\,MeV neutrino excess at reactors can be addressed with physics beyond the SM --- more precisely, via the interplay between nuclear physics and new particle physics. By injecting at least $9.4$\,MeV of energy into the nucleus, $^{13}$C may transition to the lowest-lying excited state of $^{12}$C which subsequently de-excites. The resulting $4.4$ MeV photon, together with energy deposited during neutron thermalization, can successfully explain the excess. We introduced additional, eV-scale neutrinos and assumed them to have hidden interactions with nucleons to allow them to scatter off of $^{13}$C nuclei. 

We considered three minimal types of the interactions: vector, axial and magnetic. In all cases, the parameter space preferred by Daya Bay is excluded by neutrino-induced deuterium disintegration, is in severe tension with the data collected by COHERENT and will be strongly challenged by CONUS data in the near future. While non-minimal NP scenarios could explain the reactor bump, this work suggests that the antineutrino excess in the reactor data is more likely to be of nuclear physics origin.

Interestingly, is possible to probe whether or not $^{13}$C$(\overline{\nu}, \overline{\nu}^\prime n)^{12}$C$^*$ is responsible for the reactor bump in a model-independent way. This reaction is available at all organic scintillator detectors (liquid or solid) and the prompt signal is a combination of proton recoils (about 20\%) and the 4.4\,MeV de-excitation photon. In single-volume detectors like Daya Bay and NEOS, the 4.4\,MeV photon is well contained; neither employs pulse shape discrimination, so these events pass as genuine IBD events. PROSPECT~\cite{Ashenfelter:2015uxt,Ashenfelter:2018zdm,Ashenfelter:2018iov} and STEREO~\cite{Manzanillas:2017rta,Allemandou:2018vwb,Almazan:2018wln} both rely on pulse shape discrimination to manage backgrounds, so the small proton recoil contribution will likely cause these events to be rejected as background. Also, the 4.4\,MeV photon would deposit energy across many detector cells; it seems likely that no bump would be observed in the standard analysis. The relatively long distance over which the 4.4\,MeV photon deposits its energy, compared to a positron in an IBD event, is likely to have these events rejected in finely segmented detectors like  SoLid~\cite{Abreu:2018pxg} and DANSS~\cite{Alekseev:2016llm,Alekseev:2018efk}. Curiously, DANSS seems not to observe a bump\footnote{DANSS collaboration recently announced the observation of an excess around 5 MeV with a low statistical significance. They also pointed out that more effort in the direction of calibration is required in order to robustly claim the excess (see \cite{DANSStalk} for details).}, in accordance with this prediction. A dedicated analysis in these segmented detectors could likely identify this process\footnote{Since this work first appeared on arXiv, Ref.~\cite{Zacek:2018bij} reported an observation of a bump at 5 MeV in the segmented G\"{o}sgen experiment.}.

The analogous reaction $^{17}$O$(\bar{\nu}, \bar{\nu}^\prime n)^{16}$O$^{*}$ is unlikely to be observed at water Cerenkov detectors.  First, the natural abundance of $^{17}$O ($0.03\%$) is significantly lower than that of $^{13}$C ($1.07\%$). Second, while the neutron separation energy of $^{17}$O (4143.1 keV) is slightly less than that of $^{13}$C, the lowest-lying excited state of $^{16}$O has an excitation energy of 6049.6 keV~\cite{oxygen17}; populating this state would require a significant antineutrino flux beyond $E_{\nu} \gtrsim 10$ MeV, and we have seen that this flux dies off quite quickly. Even though this could, in principle, give rise to an analogous bump in the prompt energy spectrum at $\sim6$ MeV in water Cerenkov detectors, it would be substantially more difficult to detect. Therefore, neither Gd-doped Super-K~\cite{Beacom:2003nk} nor WATCHMAN~\cite{Askins:2015bmb} should see a bump.

\textbf{Acknowledgments.} The authors would like to thank Evgeny Akhmedov, Giorgio Arcadi, Phil Barbeau, Omar Benhar, Thomas O'Donnell, Joerg Jaeckel, Joachim Kopp, Stefan Vogl and Xun-Jie Xu for useful discussions. The authors further thank Alexei Smirnov and Petr Vogel for bringing deuterium disintegration results to our attention. The work of JB and PH is supported by the U.S. Department of Energy under award number DE-SC0018327.


\appendix

\section{Deuterium disintegration cross section}
\label{app:deuterium}
The cross section for neutrino-deuterium scattering can be written \cite{Butler:1999sv,Butler:2000zp}
\begin{align}
\label{eq:DeutXSec}
\frac{d^2\sigma}{dE_f d\Omega_f} = & \, \, \frac{G_F^2 E_f^2}{2\pi^2} \times \\
& \left[ 2 W_1^{\rm LO} \sin^2 \frac{\theta_f}{2} + (W_0^{\rm LO} + W_1^{\rm LO}) \cos^2 \frac{\theta_f}{2} \right], \nonumber
\end{align}
where $E_f$ and $\Omega_f$ are the final-state neutrino energy and solid angle, respectively. The functions $W_0^{\rm LO}$ and $W_1^{\rm LO}$ are hadronic structure functions, calculated in Refs~\cite{Butler:1999sv,Butler:2000zp}. For completeness, we list the functions that enter into this cross section:
\begin{align}
W_0^{\rm LO} = & \, \, 2\left( |C_V^{(0)}|^2+|C_V^{(1)}|^2 \right) F_1 \nonumber \\
& + \left( |C_V^{(0)}|^2-|C_V^{(1)}|^2 \right) F_2 \\
& + 4 |C_V^{(0)}|^2 F^{\rm LO}_3, \nonumber \\
& \nonumber \\
W_1^{\rm LO} = & \, \, 2\left( |C_A^{(0)}|^2+|C_A^{(1)}|^2 \right) F_1 \nonumber \\
& + \frac{1}{3} \left( |C_A^{(0)}|^2-|C_A^{(1)}|^2 \right) F_2 \\
& + \frac{8}{3} |C_A^{(0)}|^2 F^{\rm LO}_3 + \frac{4}{3} |C_A^{(1)}|^2 F^{\rm LO}_4, \nonumber \\
& \nonumber \\
F_1 = & \, \, \frac{2 M_N \gamma p}{\pi (p^2+\gamma^2)^2} \left( 1 + \frac{|\vec{q}|^2(p^2-\gamma^2)}{2(p^2+\gamma^2)^2} \right), \\
F_2 = & \, \, \frac{4 M_N \gamma p}{\pi (p^2+\gamma^2)^2} \left( 1 - \frac{|\vec{q}|^2(p^2+3\gamma^2)}{6(p^2+\gamma^2)^2} \right), \\
F_3 = & \, \, \frac{1}{\pi} \Im \left[ B_0(p, |\vec{q}|)^2 A_{-1}^{(^3S_1)} \right], \\
F_4 = & \, \, \frac{1}{\pi} \Im \left[ B_0(p, |\vec{q}|)^2 A_{-1}^{(^1S_0)} \right], \\
& \nonumber \\
B_0(p, |\vec{q}|) = & \, \, - \sqrt{\frac{\gamma}{2\pi}} \frac{M_N}{\gamma-ip} \left( 1 - \frac{|\vec{q}|^2}{12(\gamma-i p)^2} \right), \\
A_{-1}^{(^3S_1)} = & \, \, -\frac{4\pi}{M_N} \frac{1}{\gamma+ip}, \\
A_{-1}^{(^1S_0)} = & \, \, -\frac{4\pi}{M_N} \frac{a^{(^1S_0)}}{1+ip \, a^{(^1S_0)}}, \\
& \nonumber \\
p = & \, \, \sqrt{M_N \nu - \gamma^2 - \frac{|\vec{q}|^2}{4}+i \epsilon} \quad (\epsilon \to 0^+), \\
|\vec{q}|^2 = & \, \, {\nu^2 + 4 E_i E_f \sin^2 \frac{\theta_f}{2}}, \\
\nu = & \, \, E_i - E_f, \\
\gamma = & \, \, \sqrt{M_N B}.
\end{align}
In these formulae, $M_N \approx 938$ MeV is the nucleon mass, $B = 2.2245$ MeV is the deuteron binding energy, $a^{(^1S_0)} = -23.7$ fm is the deuteron scattering length, $C_{V,A}^{(0)}$ is the vector/axial-vector isoscalar coupling and $C_{V,A}^{(1)}$ is the vector/axial-vector isovector coupling. The SM couplings are given by
\begin{align}
C^{(0)}_{V, \text{SM}} = & \, \, -\sin^2 \theta_W, \\
C^{(1)}_{V, \text{SM}} = & \, \, \frac{1}{2} \left(1-2\sin^2 \theta_W\right), \\
C^{(0)}_{A, \text{SM}} = & \, \, -\frac{1}{2} \Delta s, \\
C^{(1)}_{A, \text{SM}} = & \, \, \frac{1}{2} g_A,
\end{align}
where $\theta_W$ is the Weinberg angle, $\Delta s$ is the strange-quark contribution to the proton spin and $g_A$ is the $\pi N N$ coupling constant.

The new vector interaction we introduce induces additional contributions to $C_V^{(I=0,1)}$. Starting from the effective Lagrangian
\begin{align}
\mathcal{L} \supset & \, \, \frac{g_X^2 Y_\nu}{q^2-M_X^2} \left( \overline{\nu}_s \gamma^\mu \nu_s \right)\left( \overline{p} \gamma_\mu p + Y_n \overline{n} \gamma_\mu n \right),
\end{align}
we determine the NP isoscalar and isovector couplings to be
\begin{align}
\label{eq:NPC0}
C^{(0)}_{V, \text{NP}} = & \, \, \frac{\sqrt{2} g_X^2 Y_{\overline{\nu}} (1+Y_n) }{4G_F (M_X^2+4 E_i E_f \sin^2 \frac{\theta_f}{2})}, \\
C^{(1)}_{V, \text{NP}} = & \, \, \frac{-\sqrt{2} g_X^2 Y_{\overline{\nu}} (1-Y_n) }{4G_F (M_X^2+4 E_i E_f \sin^2 \frac{\theta_f}{2})},
\label{eq:NPC1}
\end{align}
where we have used that $q^2 = -4 E_i E_f \sin^2 \frac{\theta_f}{2}$. Note that a negative sign has been absorbed via $Y_{\overline{\nu}} = - Y_\nu$. We incorporate the SM and NP by adding $C^{(I)}_{V, \text{SM}}$ and $C^{(I)}_{V, \text{NP}}$, weighted by the relevant oscillation amplitude:
\begin{equation}
C_V^{(I)} = \mathcal{A}_{ee} C^{(I)}_{V, \text{SM}} + \mathcal{A}_{es} C^{(I)}_{V, \text{NP}}\,,
\end{equation}
where we need only to consider oscillations involving electron (anti)neutrinos. Squaring this and using the expressions in \cref{eq:amps,eq:amps2}, leads to
\begin{align}
|C_V^{(I)}|^2 = & \, \, P_{ee} \left(C^{(I)}_{V, \text{SM}}\right)^2 + P_{es} \left(C^{(I)}_{V, \text{NP}}\right)^2 \\
& + 2 \sin 2\theta_{ee} \cos 2\theta_{ee} \sin^2 \left( \frac{\Delta m_{41}^2 L}{4 E_{\nu}} \right) \times \nonumber \\
& \qquad \left( C^{(I)}_{V, \text{SM}} C^{(I)}_{V, \text{NP}}\right). \nonumber
\end{align}
To calculate the total cross section, \cref{eq:DeutXSec} must be integrated over the region
\begin{align}
E_f \in & \, \, \left[ 0, E_i - 2\left(M_N - \sqrt{M_N^2 - \gamma^2}\right)\right], \\
\label{eq:CosBounds}
\cos \theta_f \in & \, \, \left[ \text{Max} \left[ -1, 1-\frac{4 M_N (\nu-B)-\nu^2}{2E_i E_f} \right] ,1 \right], \\
\varphi_f \in & \, \, \left[ 0, 2\pi \right].
\end{align}

\bibliographystyle{apsrev-title}
\bibliography{refs}{}

\begin{thebibliography}{72}
\expandafter\ifx\csname natexlab\endcsname\relax\def\natexlab#1{#1}\fi
\expandafter\ifx\csname bibnamefont\endcsname\relax
  \def\bibnamefont#1{#1}\fi
\expandafter\ifx\csname bibfnamefont\endcsname\relax
  \def\bibfnamefont#1{#1}\fi
\expandafter\ifx\csname citenamefont\endcsname\relax
  \def\citenamefont#1{#1}\fi
\expandafter\ifx\csname url\endcsname\relax
  \def\url#1{\texttt{#1}}\fi
\expandafter\ifx\csname urlprefix\endcsname\relax\def\urlprefix{URL }\fi
\providecommand{\bibinfo}[2]{#2}
\providecommand{\eprint}[2][]{\url{#2}}

\bibitem[{\citenamefont{Abe et~al.}(2012)}]{Abe:2011fz}
\bibinfo{author}{\bibfnamefont{Y.}~\bibnamefont{Abe}} \bibnamefont{et~al.}
  (\bibinfo{collaboration}{Double Chooz}), ``{Indication of Reactor
  $\bar{\nu}_e$ Disappearance in the Double Chooz Experiment},''
  \bibinfo{journal}{Phys. Rev. Lett.} \textbf{\bibinfo{volume}{108}},
  \bibinfo{pages}{131801} (\bibinfo{year}{2012}), \eprint{1112.6353}.

\bibitem[{\citenamefont{An et~al.}(2012)}]{An:2012eh}
\bibinfo{author}{\bibfnamefont{F.~P.} \bibnamefont{An}} \bibnamefont{et~al.}
  (\bibinfo{collaboration}{Daya Bay}), ``{Observation of electron-antineutrino
  disappearance at Daya Bay},'' \bibinfo{journal}{Phys. Rev. Lett.}
  \textbf{\bibinfo{volume}{108}}, \bibinfo{pages}{171803}
  (\bibinfo{year}{2012}), \eprint{1203.1669}.

\bibitem[{\citenamefont{Ahn et~al.}(2012)}]{Ahn:2012nd}
\bibinfo{author}{\bibfnamefont{J.~K.} \bibnamefont{Ahn}} \bibnamefont{et~al.}
  (\bibinfo{collaboration}{RENO}), ``{Observation of Reactor Electron
  Antineutrino Disappearance in the RENO Experiment},'' \bibinfo{journal}{Phys.
  Rev. Lett.} \textbf{\bibinfo{volume}{108}}, \bibinfo{pages}{191802}
  (\bibinfo{year}{2012}), \eprint{1204.0626}.

\bibitem[{\citenamefont{Mention et~al.}(2011)\citenamefont{Mention, Fechner,
  Lasserre, Mueller, Lhuillier, Cribier, and Letourneau}}]{Mention:2011rk}
\bibinfo{author}{\bibfnamefont{G.}~\bibnamefont{Mention}},
  \bibinfo{author}{\bibfnamefont{M.}~\bibnamefont{Fechner}},
  \bibinfo{author}{\bibfnamefont{T.}~\bibnamefont{Lasserre}},
  \bibinfo{author}{\bibfnamefont{T.~A.} \bibnamefont{Mueller}},
  \bibinfo{author}{\bibfnamefont{D.}~\bibnamefont{Lhuillier}},
  \bibinfo{author}{\bibfnamefont{M.}~\bibnamefont{Cribier}}, \bibnamefont{and}
  \bibinfo{author}{\bibfnamefont{A.}~\bibnamefont{Letourneau}}, ``{The Reactor
  Antineutrino Anomaly},'' \bibinfo{journal}{Phys. Rev.}
  \textbf{\bibinfo{volume}{D83}}, \bibinfo{pages}{073006}
  (\bibinfo{year}{2011}), \eprint{1101.2755}.

\bibitem[{\citenamefont{Huber}(2016)}]{Huber:2016fkt}
\bibinfo{author}{\bibfnamefont{P.}~\bibnamefont{Huber}}, ``{Reactor
  antineutrino fluxes -- Status and challenges},'' \bibinfo{journal}{Nucl.
  Phys.} \textbf{\bibinfo{volume}{B908}}, \bibinfo{pages}{268}
  (\bibinfo{year}{2016}), \eprint{1602.01499}.

\bibitem[{\citenamefont{Huber}(2017)}]{Huber:2016xis}
\bibinfo{author}{\bibfnamefont{P.}~\bibnamefont{Huber}}, ``{NEOS Data and the
  Origin of the 5 MeV Bump in the Reactor Antineutrino Spectrum},''
  \bibinfo{journal}{Phys. Rev. Lett.} \textbf{\bibinfo{volume}{118}},
  \bibinfo{pages}{042502} (\bibinfo{year}{2017}), \eprint{1609.03910}.

\bibitem[{\citenamefont{Buck et~al.}(2017)\citenamefont{Buck, Collin, Haser,
  and Lindner}}]{Buck:2015clx}
\bibinfo{author}{\bibfnamefont{C.}~\bibnamefont{Buck}},
  \bibinfo{author}{\bibfnamefont{A.~P.} \bibnamefont{Collin}},
  \bibinfo{author}{\bibfnamefont{J.}~\bibnamefont{Haser}}, \bibnamefont{and}
  \bibinfo{author}{\bibfnamefont{M.}~\bibnamefont{Lindner}}, ``{Investigating
  the Spectral Anomaly with Different Reactor Antineutrino Experiments},''
  \bibinfo{journal}{Phys. Lett.} \textbf{\bibinfo{volume}{B765}},
  \bibinfo{pages}{159} (\bibinfo{year}{2017}), \eprint{1512.06656}.

\bibitem[{\citenamefont{Dwyer and Langford}(2015)}]{Dwyer:2014eka}
\bibinfo{author}{\bibfnamefont{D.~A.} \bibnamefont{Dwyer}} \bibnamefont{and}
  \bibinfo{author}{\bibfnamefont{T.~J.} \bibnamefont{Langford}}, ``{Spectral
  Structure of Electron Antineutrinos from Nuclear Reactors},''
  \bibinfo{journal}{Phys. Rev. Lett.} \textbf{\bibinfo{volume}{114}},
  \bibinfo{pages}{012502} (\bibinfo{year}{2015}), \eprint{1407.1281}.

\bibitem[{\citenamefont{Akimov et~al.}(2017)}]{Akimov:2017ade}
\bibinfo{author}{\bibfnamefont{D.}~\bibnamefont{Akimov}} \bibnamefont{et~al.}
  (\bibinfo{collaboration}{COHERENT}), ``{Observation of Coherent Elastic
  Neutrino-Nucleus Scattering},'' \bibinfo{journal}{Science}
  \textbf{\bibinfo{volume}{357}}, \bibinfo{pages}{1123} (\bibinfo{year}{2017}),
  \eprint{1708.01294}.

\bibitem[{\citenamefont{Reines et~al.}(1980)\citenamefont{Reines, Sobel, and
  Pasierb}}]{Reines:1980pc}
\bibinfo{author}{\bibfnamefont{F.}~\bibnamefont{Reines}},
  \bibinfo{author}{\bibfnamefont{H.~W.} \bibnamefont{Sobel}}, \bibnamefont{and}
  \bibinfo{author}{\bibfnamefont{E.}~\bibnamefont{Pasierb}}, ``{Evidence for
  Neutrino Instability},'' \bibinfo{journal}{Phys. Rev. Lett.}
  \textbf{\bibinfo{volume}{45}}, \bibinfo{pages}{1307} (\bibinfo{year}{1980}).

\bibitem[{\citenamefont{Suzuki et~al.}(1999)\citenamefont{Suzuki, Saito,
  Takahisa, Takakuwa, Nakagawa, Tohei, and Abe}}]{Suzuki:1999hr}
\bibinfo{author}{\bibfnamefont{S.}~\bibnamefont{Suzuki}},
  \bibinfo{author}{\bibfnamefont{T.}~\bibnamefont{Saito}},
  \bibinfo{author}{\bibfnamefont{K.}~\bibnamefont{Takahisa}},
  \bibinfo{author}{\bibfnamefont{C.}~\bibnamefont{Takakuwa}},
  \bibinfo{author}{\bibfnamefont{T.}~\bibnamefont{Nakagawa}},
  \bibinfo{author}{\bibfnamefont{T.}~\bibnamefont{Tohei}}, \bibnamefont{and}
  \bibinfo{author}{\bibfnamefont{K.}~\bibnamefont{Abe}}, ``{Neutron decay of
  the pygmy and giant resonances in the $^{13}\text{C} (e, e^\prime n)
  ^{12}\text{C}$ reaction},'' \bibinfo{journal}{Phys. Rev.}
  \textbf{\bibinfo{volume}{C60}}, \bibinfo{pages}{034309}
  (\bibinfo{year}{1999}).

\bibitem[{\citenamefont{An et~al.}(2017)}]{An:2016srz}
\bibinfo{author}{\bibfnamefont{F.~P.} \bibnamefont{An}} \bibnamefont{et~al.}
  (\bibinfo{collaboration}{Daya Bay}), ``{Improved Measurement of the Reactor
  Antineutrino Flux and Spectrum at Daya Bay},'' \bibinfo{journal}{Chin. Phys.}
  \textbf{\bibinfo{volume}{C41}}, \bibinfo{pages}{013002}
  (\bibinfo{year}{2017}), \eprint{1607.05378}.

\bibitem[{\citenamefont{Huber}(2011)}]{Huber:2011wv}
\bibinfo{author}{\bibfnamefont{P.}~\bibnamefont{Huber}}, ``{On the
  determination of anti-neutrino spectra from nuclear reactors},''
  \bibinfo{journal}{Phys. Rev.} \textbf{\bibinfo{volume}{C84}},
  \bibinfo{pages}{024617} (\bibinfo{year}{2011}), \bibinfo{note}{[Erratum:
  Phys. Rev.C85,029901(2012)]}, \eprint{1106.0687}.

\bibitem[{\citenamefont{Mueller et~al.}(2011)}]{Mueller:2011nm}
\bibinfo{author}{\bibfnamefont{T.~A.} \bibnamefont{Mueller}}
  \bibnamefont{et~al.}, ``{Improved Predictions of Reactor Antineutrino
  Spectra},'' \bibinfo{journal}{Phys. Rev.} \textbf{\bibinfo{volume}{C83}},
  \bibinfo{pages}{054615} (\bibinfo{year}{2011}), \eprint{1101.2663}.

\bibitem[{\citenamefont{Fallot et~al.}(2012)}]{Fallot:2012jv}
\bibinfo{author}{\bibfnamefont{M.}~\bibnamefont{Fallot}} \bibnamefont{et~al.},
  ``{New antineutrino energy spectra predictions from the summation of beta
  decay branches of the fission products},'' \bibinfo{journal}{Phys. Rev.
  Lett.} \textbf{\bibinfo{volume}{109}}, \bibinfo{pages}{202504}
  (\bibinfo{year}{2012}), \eprint{1208.3877}.

\bibitem[{\citenamefont{{Raper}}(2016)}]{NRThesis}
\bibinfo{author}{\bibfnamefont{N.}~\bibnamefont{{Raper}}}, Ph.D. thesis,
  \bibinfo{school}{Rensselaer Polytechnic Institute} (\bibinfo{year}{2016}).

\bibitem[{\citenamefont{Hayes et~al.}(2015)\citenamefont{Hayes, Friar, Garvey,
  Ibeling, Jungman, Kawano, and Mills}}]{Hayes:2015yka}
\bibinfo{author}{\bibfnamefont{A.~C.} \bibnamefont{Hayes}},
  \bibinfo{author}{\bibfnamefont{J.~L.} \bibnamefont{Friar}},
  \bibinfo{author}{\bibfnamefont{G.~T.} \bibnamefont{Garvey}},
  \bibinfo{author}{\bibfnamefont{D.}~\bibnamefont{Ibeling}},
  \bibinfo{author}{\bibfnamefont{G.}~\bibnamefont{Jungman}},
  \bibinfo{author}{\bibfnamefont{T.}~\bibnamefont{Kawano}}, \bibnamefont{and}
  \bibinfo{author}{\bibfnamefont{R.~W.} \bibnamefont{Mills}}, ``{Possible
  origins and implications of the shoulder in reactor neutrino spectra},''
  \bibinfo{journal}{Phys. Rev.} \textbf{\bibinfo{volume}{D92}},
  \bibinfo{pages}{033015} (\bibinfo{year}{2015}), \eprint{1506.00583}.

\bibitem[{\citenamefont{Birks}(1951)}]{Birks:1951boa}
\bibinfo{author}{\bibfnamefont{J.~B.} \bibnamefont{Birks}}, ``{Scintillations
  from Organic Crystals: Specific Fluorescence and Relative Response to
  Different Radiations},'' \bibinfo{journal}{Proc. Phys. Soc.}
  \textbf{\bibinfo{volume}{A64}}, \bibinfo{pages}{874} (\bibinfo{year}{1951}).

\bibitem[{\citenamefont{Minfeng}()}]{Minfeng}
\bibinfo{author}{\bibfnamefont{Y.}~\bibnamefont{Minfeng}},
  \bibinfo{note}{private communication}.

\bibitem[{pst(1993)}]{pstar}
\bibinfo{type}{Tech. Rep.} \bibinfo{number}{49},
  \bibinfo{institution}{International Commission on Radiation Units and
  Measurements} (\bibinfo{year}{1993}).

\bibitem[{\citenamefont{Ko et~al.}(2017)}]{Ko:2016owz}
\bibinfo{author}{\bibfnamefont{Y.}~\bibnamefont{Ko}} \bibnamefont{et~al.},
  ``{Sterile Neutrino Search at the NEOS Experiment},'' \bibinfo{journal}{Phys.
  Rev. Lett.} \textbf{\bibinfo{volume}{118}}, \bibinfo{pages}{121802}
  (\bibinfo{year}{2017}), \eprint{1610.05134}.

\bibitem[{\citenamefont{Maltoni and Schwetz}(2003)}]{Maltoni:2003cu}
\bibinfo{author}{\bibfnamefont{M.}~\bibnamefont{Maltoni}} \bibnamefont{and}
  \bibinfo{author}{\bibfnamefont{T.}~\bibnamefont{Schwetz}}, ``{Testing the
  statistical compatibility of independent data sets},''
  \bibinfo{journal}{Phys. Rev.} \textbf{\bibinfo{volume}{D68}},
  \bibinfo{pages}{033020} (\bibinfo{year}{2003}), \eprint{hep-ph/0304176}.

\bibitem[{\citenamefont{Nelson and Scholtz}(2011)}]{Nelson:2011sf}
\bibinfo{author}{\bibfnamefont{A.~E.} \bibnamefont{Nelson}} \bibnamefont{and}
  \bibinfo{author}{\bibfnamefont{J.}~\bibnamefont{Scholtz}}, ``{Dark Light,
  Dark Matter and the Misalignment Mechanism},'' \bibinfo{journal}{Phys. Rev.}
  \textbf{\bibinfo{volume}{D84}}, \bibinfo{pages}{103501}
  (\bibinfo{year}{2011}), \eprint{1105.2812}.

\bibitem[{\citenamefont{Buchmuller et~al.}(1991)\citenamefont{Buchmuller,
  Greub, and Minkowski}}]{Buchmuller:1991ce}
\bibinfo{author}{\bibfnamefont{W.}~\bibnamefont{Buchmuller}},
  \bibinfo{author}{\bibfnamefont{C.}~\bibnamefont{Greub}}, \bibnamefont{and}
  \bibinfo{author}{\bibfnamefont{P.}~\bibnamefont{Minkowski}}, ``{Neutrino
  masses, neutral vector bosons and the scale of $B-L$ breaking},''
  \bibinfo{journal}{Phys. Lett.} \textbf{\bibinfo{volume}{B267}},
  \bibinfo{pages}{395} (\bibinfo{year}{1991}).

\bibitem[{\citenamefont{Khalil}(2010)}]{Khalil:2010iu}
\bibinfo{author}{\bibfnamefont{S.}~\bibnamefont{Khalil}}, ``{TeV-scale gauged
  B-L symmetry with inverse seesaw mechanism},'' \bibinfo{journal}{Phys. Rev.}
  \textbf{\bibinfo{volume}{D82}}, \bibinfo{pages}{077702}
  (\bibinfo{year}{2010}), \eprint{1004.0013}.

\bibitem[{\citenamefont{Sahu and Yajnik}(2005)}]{Sahu:2004sb}
\bibinfo{author}{\bibfnamefont{N.}~\bibnamefont{Sahu}} \bibnamefont{and}
  \bibinfo{author}{\bibfnamefont{U.~A.} \bibnamefont{Yajnik}}, ``{Gauged $B -
  L$ symmetry and baryogenesis via leptogenesis at TeV scale},''
  \bibinfo{journal}{Phys. Rev.} \textbf{\bibinfo{volume}{D71}},
  \bibinfo{pages}{023507} (\bibinfo{year}{2005}), \eprint{hep-ph/0410075}.

\bibitem[{\citenamefont{Harnik et~al.}(2012)\citenamefont{Harnik, Kopp, and
  Machado}}]{Harnik:2012ni}
\bibinfo{author}{\bibfnamefont{R.}~\bibnamefont{Harnik}},
  \bibinfo{author}{\bibfnamefont{J.}~\bibnamefont{Kopp}}, \bibnamefont{and}
  \bibinfo{author}{\bibfnamefont{P.~A.~N.} \bibnamefont{Machado}}, ``{Exploring
  $\nu$ Signals in Dark Matter Detectors},'' \bibinfo{journal}{JCAP}
  \textbf{\bibinfo{volume}{1207}}, \bibinfo{pages}{026} (\bibinfo{year}{2012}),
  \eprint{1202.6073}.

\bibitem[{\citenamefont{Cerde\~{n}o et~al.}(2016)\citenamefont{Cerde\~{n}o,
  Fairbairn, Jubb, Machado, Vincent, and B{\oe}hm}}]{Cerdeno:2016sfi}
\bibinfo{author}{\bibfnamefont{D.~G.} \bibnamefont{Cerde\~{n}o}},
  \bibinfo{author}{\bibfnamefont{M.}~\bibnamefont{Fairbairn}},
  \bibinfo{author}{\bibfnamefont{T.}~\bibnamefont{Jubb}},
  \bibinfo{author}{\bibfnamefont{P.~A.~N.} \bibnamefont{Machado}},
  \bibinfo{author}{\bibfnamefont{A.~C.} \bibnamefont{Vincent}},
  \bibnamefont{and} \bibinfo{author}{\bibfnamefont{C.}~\bibnamefont{B{\oe}hm}},
  ``{Physics from solar neutrinos in dark matter direct detection
  experiments},'' \bibinfo{journal}{JHEP} \textbf{\bibinfo{volume}{05}},
  \bibinfo{pages}{118} (\bibinfo{year}{2016}), \bibinfo{note}{[Erratum:
  JHEP09,048(2016)]}, \eprint{1604.01025}.

\bibitem[{\citenamefont{Pospelov}(2011)}]{Pospelov:2011ha}
\bibinfo{author}{\bibfnamefont{M.}~\bibnamefont{Pospelov}}, ``{Neutrino Physics
  with Dark Matter Experiments and the Signature of New Baryonic Neutral
  Currents},'' \bibinfo{journal}{Phys. Rev.} \textbf{\bibinfo{volume}{D84}},
  \bibinfo{pages}{085008} (\bibinfo{year}{2011}), \eprint{1103.3261}.

\bibitem[{\citenamefont{Kopp and Welter}(2014)}]{Kopp:2014fha}
\bibinfo{author}{\bibfnamefont{J.}~\bibnamefont{Kopp}} \bibnamefont{and}
  \bibinfo{author}{\bibfnamefont{J.}~\bibnamefont{Welter}}, ``{The
  Not-So-Sterile 4th Neutrino: Constraints on New Gauge Interactions from
  Neutrino Oscillation Experiments},'' \bibinfo{journal}{JHEP}
  \textbf{\bibinfo{volume}{12}}, \bibinfo{pages}{104} (\bibinfo{year}{2014}),
  \eprint{1408.0289}.

\bibitem[{\citenamefont{Kopp et~al.}(2013)\citenamefont{Kopp, Machado, Maltoni,
  and Schwetz}}]{Kopp:2013vaa}
\bibinfo{author}{\bibfnamefont{J.}~\bibnamefont{Kopp}},
  \bibinfo{author}{\bibfnamefont{P.~A.~N.} \bibnamefont{Machado}},
  \bibinfo{author}{\bibfnamefont{M.}~\bibnamefont{Maltoni}}, \bibnamefont{and}
  \bibinfo{author}{\bibfnamefont{T.}~\bibnamefont{Schwetz}}, ``{Sterile
  Neutrino Oscillations: The Global Picture},'' \bibinfo{journal}{JHEP}
  \textbf{\bibinfo{volume}{05}}, \bibinfo{pages}{050} (\bibinfo{year}{2013}),
  \eprint{1303.3011}.

\bibitem[{\citenamefont{Gariazzo et~al.}(2017)\citenamefont{Gariazzo, Giunti,
  Laveder, and Li}}]{Gariazzo:2017fdh}
\bibinfo{author}{\bibfnamefont{S.}~\bibnamefont{Gariazzo}},
  \bibinfo{author}{\bibfnamefont{C.}~\bibnamefont{Giunti}},
  \bibinfo{author}{\bibfnamefont{M.}~\bibnamefont{Laveder}}, \bibnamefont{and}
  \bibinfo{author}{\bibfnamefont{Y.~F.} \bibnamefont{Li}}, ``{Updated Global
  3+1 Analysis of Short-BaseLine Neutrino Oscillations},''
  \bibinfo{journal}{JHEP} \textbf{\bibinfo{volume}{06}}, \bibinfo{pages}{135}
  (\bibinfo{year}{2017}), \eprint{1703.00860}.

\bibitem[{\citenamefont{Dentler et~al.}(2017)\citenamefont{Dentler,
  Hern\'{a}ndez-Cabezudo, Kopp, Maltoni, and Schwetz}}]{Dentler:2017tkw}
\bibinfo{author}{\bibfnamefont{M.}~\bibnamefont{Dentler}},
  \bibinfo{author}{\bibfnamefont{A.}~\bibnamefont{Hern\'{a}ndez-Cabezudo}},
  \bibinfo{author}{\bibfnamefont{J.}~\bibnamefont{Kopp}},
  \bibinfo{author}{\bibfnamefont{M.}~\bibnamefont{Maltoni}}, \bibnamefont{and}
  \bibinfo{author}{\bibfnamefont{T.}~\bibnamefont{Schwetz}}, ``{Sterile
  neutrinos or flux uncertainties? --- Status of the reactor anti-neutrino
  anomaly},'' \bibinfo{journal}{JHEP} \textbf{\bibinfo{volume}{11}},
  \bibinfo{pages}{099} (\bibinfo{year}{2017}), \eprint{1709.04294}.

\bibitem[{\citenamefont{Gariazzo et~al.}(2018)\citenamefont{Gariazzo, Giunti,
  Laveder, and Li}}]{Gariazzo:2018mwd}
\bibinfo{author}{\bibfnamefont{S.}~\bibnamefont{Gariazzo}},
  \bibinfo{author}{\bibfnamefont{C.}~\bibnamefont{Giunti}},
  \bibinfo{author}{\bibfnamefont{M.}~\bibnamefont{Laveder}}, \bibnamefont{and}
  \bibinfo{author}{\bibfnamefont{Y.~F.} \bibnamefont{Li}}, ``{Model-independent
  $\bar\nu_{e}$ short-baseline oscillations from reactor spectral ratios},''
  \bibinfo{journal}{Phys. Lett.} \textbf{\bibinfo{volume}{B782}},
  \bibinfo{pages}{13} (\bibinfo{year}{2018}), \eprint{1801.06467}.

\bibitem[{\citenamefont{Dentler et~al.}(2018)\citenamefont{Dentler,
  Hernández-Cabezudo, Kopp, Machado, Maltoni, Martinez-Soler, and
  Schwetz}}]{Dentler:2018sju}
\bibinfo{author}{\bibfnamefont{M.}~\bibnamefont{Dentler}},
  \bibinfo{author}{\bibfnamefont{Ã.}~\bibnamefont{Hernández-Cabezudo}},
  \bibinfo{author}{\bibfnamefont{J.}~\bibnamefont{Kopp}},
  \bibinfo{author}{\bibfnamefont{P.~A.~N.} \bibnamefont{Machado}},
  \bibinfo{author}{\bibfnamefont{M.}~\bibnamefont{Maltoni}},
  \bibinfo{author}{\bibfnamefont{I.}~\bibnamefont{Martinez-Soler}},
  \bibnamefont{and} \bibinfo{author}{\bibfnamefont{T.}~\bibnamefont{Schwetz}},
  ``{Updated Global Analysis of Neutrino Oscillations in the Presence of
  eV-Scale Sterile Neutrinos},'' \bibinfo{journal}{JHEP}
  \textbf{\bibinfo{volume}{08}}, \bibinfo{pages}{010} (\bibinfo{year}{2018}),
  \eprint{1803.10661}.

\bibitem[{\citenamefont{Atiya et~al.}(1992)}]{Atiya:1992sm}
\bibinfo{author}{\bibfnamefont{M.~S.} \bibnamefont{Atiya}}
  \bibnamefont{et~al.}, ``{Search for the decay $\pi^0 \to \gamma + X$},''
  \bibinfo{journal}{Phys. Rev. Lett.} \textbf{\bibinfo{volume}{69}},
  \bibinfo{pages}{733} (\bibinfo{year}{1992}).

\bibitem[{\citenamefont{Batell et~al.}(2014)\citenamefont{Batell, deNiverville,
  McKeen, Pospelov, and Ritz}}]{Batell:2014yra}
\bibinfo{author}{\bibfnamefont{B.}~\bibnamefont{Batell}},
  \bibinfo{author}{\bibfnamefont{P.}~\bibnamefont{deNiverville}},
  \bibinfo{author}{\bibfnamefont{D.}~\bibnamefont{McKeen}},
  \bibinfo{author}{\bibfnamefont{M.}~\bibnamefont{Pospelov}}, \bibnamefont{and}
  \bibinfo{author}{\bibfnamefont{A.}~\bibnamefont{Ritz}}, ``{Leptophobic Dark
  Matter at Neutrino Factories},'' \bibinfo{journal}{Phys. Rev.}
  \textbf{\bibinfo{volume}{D90}}, \bibinfo{pages}{115014}
  (\bibinfo{year}{2014}), \eprint{1405.7049}.

\bibitem[{\citenamefont{deNiverville et~al.}(2015)\citenamefont{deNiverville,
  Pospelov, and Ritz}}]{deNiverville:2015mwa}
\bibinfo{author}{\bibfnamefont{P.}~\bibnamefont{deNiverville}},
  \bibinfo{author}{\bibfnamefont{M.}~\bibnamefont{Pospelov}}, \bibnamefont{and}
  \bibinfo{author}{\bibfnamefont{A.}~\bibnamefont{Ritz}}, ``{Light new physics
  in coherent neutrino-nucleus scattering experiments},''
  \bibinfo{journal}{Phys. Rev.} \textbf{\bibinfo{volume}{D92}},
  \bibinfo{pages}{095005} (\bibinfo{year}{2015}), \eprint{1505.07805}.

\bibitem[{\citenamefont{Barbieri and Ericson}(1975)}]{Barbieri:1975xy}
\bibinfo{author}{\bibfnamefont{R.}~\bibnamefont{Barbieri}} \bibnamefont{and}
  \bibinfo{author}{\bibfnamefont{T.~E.~O.} \bibnamefont{Ericson}}, ``{Evidence
  Against the Existence of a Low Mass Scalar Boson from Neutron-Nucleus
  Scattering},'' \bibinfo{journal}{Phys. Lett.} \textbf{\bibinfo{volume}{57B}},
  \bibinfo{pages}{270} (\bibinfo{year}{1975}).

\bibitem[{\citenamefont{Leeb and Schmiedmayer}(1992)}]{Leeb:1992qf}
\bibinfo{author}{\bibfnamefont{H.}~\bibnamefont{Leeb}} \bibnamefont{and}
  \bibinfo{author}{\bibfnamefont{J.}~\bibnamefont{Schmiedmayer}}, ``{Constraint
  on hypothetical light interacting bosons from low-energy neutron
  experiments},'' \bibinfo{journal}{Phys. Rev. Lett.}
  \textbf{\bibinfo{volume}{68}}, \bibinfo{pages}{1472} (\bibinfo{year}{1992}).

\bibitem[{\citenamefont{Ecker et~al.}(1987)\citenamefont{Ecker, Pich, and
  de~Rafael}}]{Ecker:1987qi}
\bibinfo{author}{\bibfnamefont{G.}~\bibnamefont{Ecker}},
  \bibinfo{author}{\bibfnamefont{A.}~\bibnamefont{Pich}}, \bibnamefont{and}
  \bibinfo{author}{\bibfnamefont{E.}~\bibnamefont{de~Rafael}}, ``{$K \to \pi
  \ell^+ \ell^-$ Decays in the Effective Chiral Lagrangian of the Standard
  Model},'' \bibinfo{journal}{Nucl. Phys.} \textbf{\bibinfo{volume}{B291}},
  \bibinfo{pages}{692} (\bibinfo{year}{1987}).

\bibitem[{\citenamefont{Ecker et~al.}(1988)\citenamefont{Ecker, Pich, and
  de~Rafael}}]{Ecker:1987hd}
\bibinfo{author}{\bibfnamefont{G.}~\bibnamefont{Ecker}},
  \bibinfo{author}{\bibfnamefont{A.}~\bibnamefont{Pich}}, \bibnamefont{and}
  \bibinfo{author}{\bibfnamefont{E.}~\bibnamefont{de~Rafael}}, ``{Radiative
  Kaon Decays and CP Violation in Chiral Perturbation Theory},''
  \bibinfo{journal}{Nucl. Phys.} \textbf{\bibinfo{volume}{B303}},
  \bibinfo{pages}{665} (\bibinfo{year}{1988}).

\bibitem[{\citenamefont{D'Ambrosio et~al.}(1998)\citenamefont{D'Ambrosio,
  Ecker, Isidori, and Portoles}}]{DAmbrosio:1998gur}
\bibinfo{author}{\bibfnamefont{G.}~\bibnamefont{D'Ambrosio}},
  \bibinfo{author}{\bibfnamefont{G.}~\bibnamefont{Ecker}},
  \bibinfo{author}{\bibfnamefont{G.}~\bibnamefont{Isidori}}, \bibnamefont{and}
  \bibinfo{author}{\bibfnamefont{J.}~\bibnamefont{Portoles}}, ``{The Decays $K
  \to \pi \ell^+ \ell^-$ beyond leading order in the chiral expansion},''
  \bibinfo{journal}{JHEP} \textbf{\bibinfo{volume}{08}}, \bibinfo{pages}{004}
  (\bibinfo{year}{1998}), \eprint{hep-ph/9808289}.

\bibitem[{\citenamefont{Pospelov}(2009)}]{Pospelov:2008zw}
\bibinfo{author}{\bibfnamefont{M.}~\bibnamefont{Pospelov}}, ``{Secluded U(1)
  below the weak scale},'' \bibinfo{journal}{Phys. Rev.}
  \textbf{\bibinfo{volume}{D80}}, \bibinfo{pages}{095002}
  (\bibinfo{year}{2009}), \eprint{0811.1030}.

\bibitem[{\citenamefont{Maneschg}()}]{CONUStalk}
\bibinfo{author}{\bibfnamefont{W.}~\bibnamefont{Maneschg}}, ``{The status of
  CONUS},'', \urlprefix\url{http://doi.org/10.5281/zenodo.1286927}.

\bibitem[{\citenamefont{Klein and Nystrand}(2000)}]{Klein:1999gv}
\bibinfo{author}{\bibfnamefont{S.~R.} \bibnamefont{Klein}} \bibnamefont{and}
  \bibinfo{author}{\bibfnamefont{J.}~\bibnamefont{Nystrand}}, ``{Interference
  in exclusive vector meson production in heavy ion collisions},''
  \bibinfo{journal}{Phys. Rev. Lett.} \textbf{\bibinfo{volume}{84}},
  \bibinfo{pages}{2330} (\bibinfo{year}{2000}), \eprint{hep-ph/9909237}.

\bibitem[{\citenamefont{Liao and Marfatia}(2017)}]{Liao:2017uzy}
\bibinfo{author}{\bibfnamefont{J.}~\bibnamefont{Liao}} \bibnamefont{and}
  \bibinfo{author}{\bibfnamefont{D.}~\bibnamefont{Marfatia}}, ``{COHERENT
  constraints on nonstandard neutrino interactions},'' \bibinfo{journal}{Phys.
  Lett.} \textbf{\bibinfo{volume}{B775}}, \bibinfo{pages}{54}
  (\bibinfo{year}{2017}), \eprint{1708.04255}.

\bibitem[{\citenamefont{Farzan et~al.}(2018)\citenamefont{Farzan, Lindner,
  Rodejohann, and Xu}}]{Farzan:2018gtr}
\bibinfo{author}{\bibfnamefont{Y.}~\bibnamefont{Farzan}},
  \bibinfo{author}{\bibfnamefont{M.}~\bibnamefont{Lindner}},
  \bibinfo{author}{\bibfnamefont{W.}~\bibnamefont{Rodejohann}},
  \bibnamefont{and} \bibinfo{author}{\bibfnamefont{X.-J.} \bibnamefont{Xu}},
  ``{Probing neutrino coupling to a light scalar with coherent neutrino
  scattering},'' \bibinfo{journal}{JHEP} \textbf{\bibinfo{volume}{05}},
  \bibinfo{pages}{066} (\bibinfo{year}{2018}), \eprint{1802.05171}.

\bibitem[{\citenamefont{Butler and Chen}(2000)}]{Butler:1999sv}
\bibinfo{author}{\bibfnamefont{M.}~\bibnamefont{Butler}} \bibnamefont{and}
  \bibinfo{author}{\bibfnamefont{J.-W.} \bibnamefont{Chen}}, ``{Elastic and
  inelastic neutrino deuteron scattering in effective field theory},''
  \bibinfo{journal}{Nucl. Phys.} \textbf{\bibinfo{volume}{A675}},
  \bibinfo{pages}{575} (\bibinfo{year}{2000}), \eprint{nucl-th/9905059}.

\bibitem[{\citenamefont{Butler et~al.}(2001)\citenamefont{Butler, Chen, and
  Kong}}]{Butler:2000zp}
\bibinfo{author}{\bibfnamefont{M.}~\bibnamefont{Butler}},
  \bibinfo{author}{\bibfnamefont{J.-W.} \bibnamefont{Chen}}, \bibnamefont{and}
  \bibinfo{author}{\bibfnamefont{X.}~\bibnamefont{Kong}}, ``{Neutrino deuteron
  scattering in effective field theory at next-to-next-to-leading order},''
  \bibinfo{journal}{Phys. Rev.} \textbf{\bibinfo{volume}{C63}},
  \bibinfo{pages}{035501} (\bibinfo{year}{2001}), \eprint{nucl-th/0008032}.

\bibitem[{\citenamefont{Barranco et~al.}(2005)\citenamefont{Barranco, Miranda,
  and Rashba}}]{Barranco:2005yy}
\bibinfo{author}{\bibfnamefont{J.}~\bibnamefont{Barranco}},
  \bibinfo{author}{\bibfnamefont{O.~G.} \bibnamefont{Miranda}},
  \bibnamefont{and} \bibinfo{author}{\bibfnamefont{T.~I.}
  \bibnamefont{Rashba}}, ``{Probing new physics with coherent neutrino
  scattering off nuclei},'' \bibinfo{journal}{JHEP}
  \textbf{\bibinfo{volume}{12}}, \bibinfo{pages}{021} (\bibinfo{year}{2005}),
  \eprint{hep-ph/0508299}.

\bibitem[{\citenamefont{Scholberg}(2018)}]{Scholberg:2018vwg}
\bibinfo{author}{\bibfnamefont{K.}~\bibnamefont{Scholberg}}
  (\bibinfo{collaboration}{COHERENT}), in \emph{\bibinfo{booktitle}{{2017
  International Workshop on Neutrinos from Accelerators (NuFact17) Uppsala
  University Main Building, Uppsala, Sweden, September 25-30, 2017}}}
  (\bibinfo{year}{2018}), \eprint{1801.05546},
  \urlprefix\url{https://inspirehep.net/record/1648558/files/arXiv:1801.05546.pdf}.

\bibitem[{\citenamefont{Grifols et~al.}(2004)\citenamefont{Grifols, Masso, and
  Mohanty}}]{Grifols:2004yn}
\bibinfo{author}{\bibfnamefont{J.~A.} \bibnamefont{Grifols}},
  \bibinfo{author}{\bibfnamefont{E.}~\bibnamefont{Masso}}, \bibnamefont{and}
  \bibinfo{author}{\bibfnamefont{S.}~\bibnamefont{Mohanty}}, ``{Neutrino
  magnetic moments and photodisintegration of deuterium},''
  \bibinfo{journal}{Phys. Lett.} \textbf{\bibinfo{volume}{B587}},
  \bibinfo{pages}{184} (\bibinfo{year}{2004}), \eprint{hep-ph/0401144}.

\bibitem[{\citenamefont{Koning and Rochman}(2012)}]{Koning:2012zqy}
\bibinfo{author}{\bibfnamefont{A.~J.} \bibnamefont{Koning}} \bibnamefont{and}
  \bibinfo{author}{\bibfnamefont{D.}~\bibnamefont{Rochman}}, ``{Modern Nuclear
  Data Evaluation with the TALYS Code System},'' \bibinfo{journal}{Nucl. Data
  Sheets} \textbf{\bibinfo{volume}{113}}, \bibinfo{pages}{2841}
  (\bibinfo{year}{2012}).

\bibitem[{\citenamefont{Fitzpatrick et~al.}(2013)\citenamefont{Fitzpatrick,
  Haxton, Katz, Lubbers, and Xu}}]{Fitzpatrick:2012ix}
\bibinfo{author}{\bibfnamefont{A.~L.} \bibnamefont{Fitzpatrick}},
  \bibinfo{author}{\bibfnamefont{W.}~\bibnamefont{Haxton}},
  \bibinfo{author}{\bibfnamefont{E.}~\bibnamefont{Katz}},
  \bibinfo{author}{\bibfnamefont{N.}~\bibnamefont{Lubbers}}, \bibnamefont{and}
  \bibinfo{author}{\bibfnamefont{Y.}~\bibnamefont{Xu}}, ``{The Effective Field
  Theory of Dark Matter Direct Detection},'' \bibinfo{journal}{JCAP}
  \textbf{\bibinfo{volume}{1302}}, \bibinfo{pages}{004} (\bibinfo{year}{2013}),
  \eprint{1203.3542}.

\bibitem[{\citenamefont{Anand et~al.}(2014)\citenamefont{Anand, Fitzpatrick,
  and Haxton}}]{Anand:2013yka}
\bibinfo{author}{\bibfnamefont{N.}~\bibnamefont{Anand}},
  \bibinfo{author}{\bibfnamefont{A.~L.} \bibnamefont{Fitzpatrick}},
  \bibnamefont{and} \bibinfo{author}{\bibfnamefont{W.~C.}
  \bibnamefont{Haxton}}, ``{Weakly interacting massive particle-nucleus elastic
  scattering response},'' \bibinfo{journal}{Phys. Rev.}
  \textbf{\bibinfo{volume}{C89}}, \bibinfo{pages}{065501}
  (\bibinfo{year}{2014}), \eprint{1308.6288}.

\bibitem[{\citenamefont{Berryman}()}]{upcoming}
\bibinfo{author}{\bibfnamefont{J.~M.} \bibnamefont{Berryman}},
  \bibinfo{note}{in preparation}.

\bibitem[{\citenamefont{Akhmedov and Berezin}(1992)}]{Akhmedov:1991rt}
\bibinfo{author}{\bibfnamefont{E.~K.} \bibnamefont{Akhmedov}} \bibnamefont{and}
  \bibinfo{author}{\bibfnamefont{V.~V.} \bibnamefont{Berezin}}, ``{Neutrino
  disintegration of deuteron and electromagnetic form-factors of neutrinos},''
  \bibinfo{journal}{Z. Phys.} \textbf{\bibinfo{volume}{C54}},
  \bibinfo{pages}{661} (\bibinfo{year}{1992}).

\bibitem[{\citenamefont{Ashenfelter et~al.}(2016)}]{Ashenfelter:2015uxt}
\bibinfo{author}{\bibfnamefont{J.}~\bibnamefont{Ashenfelter}}
  \bibnamefont{et~al.} (\bibinfo{collaboration}{PROSPECT}), ``{The PROSPECT
  Physics Program},'' \bibinfo{journal}{J. Phys.}
  \textbf{\bibinfo{volume}{G43}}, \bibinfo{pages}{113001}
  (\bibinfo{year}{2016}), \eprint{1512.02202}.

\bibitem[{\citenamefont{Ashenfelter et~al.}(2019)}]{Ashenfelter:2018zdm}
\bibinfo{author}{\bibfnamefont{J.}~\bibnamefont{Ashenfelter}}
  \bibnamefont{et~al.} (\bibinfo{collaboration}{PROSPECT}), ``{The PROSPECT
  Reactor Antineutrino Experiment},'' \bibinfo{journal}{Nucl. Instrum. Meth.}
  \textbf{\bibinfo{volume}{A922}}, \bibinfo{pages}{287} (\bibinfo{year}{2019}),
  \eprint{1808.00097}.

\bibitem[{\citenamefont{Ashenfelter et~al.}(2018)}]{Ashenfelter:2018iov}
\bibinfo{author}{\bibfnamefont{J.}~\bibnamefont{Ashenfelter}}
  \bibnamefont{et~al.} (\bibinfo{collaboration}{PROSPECT}), ``{First search for
  short-baseline neutrino oscillations at HFIR with PROSPECT},''
  \bibinfo{journal}{Phys. Rev. Lett.} \textbf{\bibinfo{volume}{121}},
  \bibinfo{pages}{251802} (\bibinfo{year}{2018}), \eprint{1806.02784}.

\bibitem[{\citenamefont{Manzanillas}(2017)}]{Manzanillas:2017rta}
\bibinfo{author}{\bibfnamefont{L.}~\bibnamefont{Manzanillas}}
  (\bibinfo{collaboration}{STEREO}), ``{STEREO: Search for sterile neutrinos at
  the ILL},'' \bibinfo{journal}{PoS} \textbf{\bibinfo{volume}{NOW2016}},
  \bibinfo{pages}{033} (\bibinfo{year}{2017}), \eprint{1702.02498}.

\bibitem[{\citenamefont{Allemandou et~al.}(2018)}]{Allemandou:2018vwb}
\bibinfo{author}{\bibfnamefont{N.}~\bibnamefont{Allemandou}}
  \bibnamefont{et~al.} (\bibinfo{collaboration}{STEREO}), ``{The STEREO
  Experiment},'' \bibinfo{journal}{JINST} \textbf{\bibinfo{volume}{13}},
  \bibinfo{pages}{P07009} (\bibinfo{year}{2018}), \eprint{1804.09052}.

\bibitem[{\citenamefont{Almazán et~al.}(2018)}]{Almazan:2018wln}
\bibinfo{author}{\bibfnamefont{H.}~\bibnamefont{Almazán}} \bibnamefont{et~al.}
  (\bibinfo{collaboration}{STEREO}), ``{Sterile Neutrino Constraints from the
  STEREO Experiment with 66 Days of Reactor-On Data},'' \bibinfo{journal}{Phys.
  Rev. Lett.} \textbf{\bibinfo{volume}{121}}, \bibinfo{pages}{161801}
  (\bibinfo{year}{2018}), \eprint{1806.02096}.

\bibitem[{\citenamefont{Abreu et~al.}(2018)}]{Abreu:2018pxg}
\bibinfo{author}{\bibfnamefont{Y.}~\bibnamefont{Abreu}} \bibnamefont{et~al.}
  (\bibinfo{collaboration}{SoLid}), ``{Performance of a full scale prototype
  detector at the BR2 reactor for the SoLid experiment},''
  \bibinfo{journal}{Submitted to: JINST}  (\bibinfo{year}{2018}),
  \eprint{1802.02884}.

\bibitem[{\citenamefont{Alekseev et~al.}(2016)}]{Alekseev:2016llm}
\bibinfo{author}{\bibfnamefont{I.}~\bibnamefont{Alekseev}}
  \bibnamefont{et~al.}, ``{DANSS: Detector of the reactor AntiNeutrino based on
  Solid Scintillator},'' \bibinfo{journal}{JINST}
  \textbf{\bibinfo{volume}{11}}, \bibinfo{pages}{P11011}
  (\bibinfo{year}{2016}), \eprint{1606.02896}.

\bibitem[{\citenamefont{Alekseev et~al.}(2018)}]{Alekseev:2018efk}
\bibinfo{author}{\bibfnamefont{I.}~\bibnamefont{Alekseev}} \bibnamefont{et~al.}
  (\bibinfo{collaboration}{DANSS}), ``{Search for sterile neutrinos at the
  DANSS experiment},'' \bibinfo{journal}{Phys. Lett.}
  \textbf{\bibinfo{volume}{B787}}, \bibinfo{pages}{56} (\bibinfo{year}{2018}),
  \eprint{1804.04046}.

\bibitem[{\citenamefont{Svirida}()}]{DANSStalk}
\bibinfo{author}{\bibfnamefont{D.}~\bibnamefont{Svirida}}, ``{Searches for
  sterile neutrinos at the DANSSexperiment},'',
  \urlprefix\url{http://www.ba.infn.it/~now/now2018/assets/svirida_danss_now18.pdf}.

\bibitem[{\citenamefont{Zacek et~al.}(2018)\citenamefont{Zacek, Zacek, Vogel,
  and Vuilleumier}}]{Zacek:2018bij}
\bibinfo{author}{\bibfnamefont{V.}~\bibnamefont{Zacek}},
  \bibinfo{author}{\bibfnamefont{G.}~\bibnamefont{Zacek}},
  \bibinfo{author}{\bibfnamefont{P.}~\bibnamefont{Vogel}}, \bibnamefont{and}
  \bibinfo{author}{\bibfnamefont{J.~L.} \bibnamefont{Vuilleumier}}, ``{Evidence
  for a 5 MeV Spectral Deviation in the Goesgen Reactor Neutrino Oscillation
  Experiment},''  (\bibinfo{year}{2018}), \eprint{1807.01810}.

\bibitem[{\citenamefont{J.H.~Kelley and Cheves}(1993)}]{oxygen17}
\bibinfo{author}{\bibfnamefont{H.~W.} \bibnamefont{J.H.~Kelley},
  \bibfnamefont{D.R.~Tilley}} \bibnamefont{and}
  \bibinfo{author}{\bibfnamefont{C.}~\bibnamefont{Cheves}}
  \bibinfo{journal}{Nucl. Physics} \textbf{\bibinfo{volume}{564}},
  \bibinfo{pages}{1} (\bibinfo{year}{1993}).

\bibitem[{\citenamefont{Beacom and Vagins}(2004)}]{Beacom:2003nk}
\bibinfo{author}{\bibfnamefont{J.~F.} \bibnamefont{Beacom}} \bibnamefont{and}
  \bibinfo{author}{\bibfnamefont{M.~R.} \bibnamefont{Vagins}}, ``{GADZOOKS!
  Anti-neutrino spectroscopy with large water Cherenkov detectors},''
  \bibinfo{journal}{Phys. Rev. Lett.} \textbf{\bibinfo{volume}{93}},
  \bibinfo{pages}{171101} (\bibinfo{year}{2004}), \eprint{hep-ph/0309300}.

\bibitem[{\citenamefont{Askins et~al.}(2015)}]{Askins:2015bmb}
\bibinfo{author}{\bibfnamefont{M.}~\bibnamefont{Askins}} \bibnamefont{et~al.}
  (\bibinfo{collaboration}{WATCHMAN}), ``{The Physics and Nuclear
  Nonproliferation Goals of WATCHMAN: A WAter CHerenkov Monitor for
  ANtineutrinos},''  (\bibinfo{year}{2015}), \eprint{1502.01132}.

\end{thebibliography}

\end{document}